\documentclass{emulateapj}
\usepackage{graphicx}
\newcommand{\secpoint}{\mbox{$''\mskip-7.6mu.\,$}}
\newcommand{\minpoint}{\mbox{$\,'\mskip-7.6mu.\,$}}

\slugcomment{accepted March 18, 2016}

\begin{document}

\title{A Tale of Two Narrow-Line Regions: Ionization, Kinematics, and Spectral Energy Distributions for a Local Pair of Merging Obscured Active Galaxies}

\shorttitle{Ionization and Kinematics for a Local Pair of Obscured AGNs}
\shortauthors{HAINLINE ET AL.}

\author{\sc Kevin N. Hainline}
\affil{Steward Observatory, 933 N. Cherry St., University of Arizona, Tucson, AZ 85721}

\author{Ryan C. Hickox, Chien-Ting Chen, Christopher M. Carroll, Mackenzie L. Jones, Alexandros S. Zervos}
\affil{Department of Physics and Astronomy, Dartmouth College, Hanover, NH 03755}

\author{Andrew D. Goulding}
\affil{Department Astrophysical Sciences, Princeton University, Princeton, NJ 08544}

\begin{abstract}

We explore the gas ionization and kinematics, as well as the optical--IR spectral energy distributions for UGC 11185, a nearby pair of merging galaxies hosting obscured active galactic nuclei (AGNs), also known as SDSS J181611.72+423941.6 and J181609.37+423923.0 (J1816NE and J1816SW, $z \approx 0.04$). Due to the wide separation between these interacting galaxies ($\sim 23$ kpc), observations of these objects provide a rare glimpse of the concurrent growth of supermassive black holes at an early merger stage. We use BPT line diagnostics to show that the full extent of the narrow line emission in both galaxies is photoionized by an AGN and confirm the existence of a 10-kpc-scale ionization cone in J1816NE, while in J1816SW the AGN narrow-line region is much more compact (1--2 kpc) and relatively undisturbed. Our observations also reveal the presence of ionized gas that nearly spans the entire distance between the galaxies which is likely in a merger-induced tidal stream. In addition, we carry out a spectral analysis of the X-ray emission using data from {\em XMM-Newton}. These galaxies represent a useful pair to explore how the [\ion{O}{3}] luminosity of an AGN is dependent on the size of the region used to explore the extended emission. Given the growing evidence for AGN ``flickering'' over short timescales, we speculate that the appearances and impact of these AGNs may change multiple times over the course of the galaxy merger, which is especially important given that these objects are likely the progenitors of the types of systems commonly classified as ``dual AGNs.'' 

\end{abstract}


\section{Introduction}
\label{sec:intro}

A great deal of insight into the history, luminosity and kinematic influence of an Active Galactic Nucleus (AGN) can be found by exploring the extended emission line regions found around many of these objects. In these kpc-scale narrow-line regions (NLRs) gas is thought to be photoionized by the powerful emission from the accretion disk very near the central supermassive black hole. While extended NLRs are ubiquitous in luminous quasars \citep{wampler1975,stockton1976,bennert2002,greene2011,liu2013,hainline2013,hainline2014a}, there are a subset of lower-luminosity objects with extended regions which have been used to understand AGN luminosity history as well as AGN ionization geometry \citep[][]{lintott2009, schawinski2010, keel2012, schirmer2013, keel2015}. These objects are often found in mergers, as galaxy interactions are thought to be a driver of gas towards the centers of galaxies \citep{hopkins2008}. In fact, mergers have been shown to lead to an increase in AGN activity \citep[e.g.,][]{koss2010, ellison2011,bessiere2012, sabater2013} while also leading to an extended ejected gas distribution which can be photoionized by an AGN \citep{keel2012}. Detailed numerical simulations of galaxy mergers, which occur on timescales of billions of years, have found that black hole accretion increases in the first Gyr after the two galaxies experienced their first encounter, and subsequently peaks during coalescence \citep[e.g.,][]{dimatteo2005,cox2008,hopkins2008}.

The association of AGN activity with galaxy mergers suggests the existence of ``binary'' or ``dual'' AGN, in which both black holes are active during the process of merging and coalescence. Such systems have now been observed over a range of black hole separations, although the number of confirmed detections that have been studied in detail remains relatively small \citep[e.g.][and references therein]{comerford2015, komossa2003, koss2011, mazzarella2012}.

The interpretation of binary or dual AGN is made more complex by the fact that AGN activity can vary significantly on timescales of millions of years or less \citep[e.g.,][]{hickox2014,schawinski2015}, and an infalling pair of merging galaxies may transition from active and inactive states many times during the lifetime of the merger. This poses a problem for understanding any potential correlations between AGN activity and galaxy mergers in large surveys, increasing the importance of finding mergers where both galaxies are observed to be active for relating AGN activity and merger state. Moreover, major mergers are theorized to eventually lead to the creation of powerful obscured quasars \citep[e.g.,][]{hopkins2006}, and studies of the AGN properties of galaxies in earlier merger stages offers an opportunity to trace the evolution of these objects. 

In addition, AGNs with extended emission regions can also be used to explore AGN feedback, where the effects of powerful AGNs can drive gas from galaxies and quench star formation and fueling of the central black hole. Powerful, high-velocity winds are observed in many galaxies \citep{heckman2000, veilleux2005, tremonti2007, weiner2009, hainline2011}, but disentangling the effects of the AGN and stellar processes within galaxies can be difficult \citep[see][for a review]{fabian2012}. If AGN activity is triggered by mergers, we can use detailed observations of local mergers to measure the ionization properties of excited, extended NLR gas, which can then be used along with gas kinematics to explore the nature of AGN feedback in merging galaxies.

The extended NLR is also of particular interest for understanding AGN luminosity, one of the fundamental properties of an AGN that is used to classify these objects. There are many indicators of AGN luminosity \citep[see][]{diamondstanic2009,lamassa2010}, including the flux of the [\ion{O}{3}]$\lambda$5007 emission line, a strong feature that is seen in AGN NLRs \citep{bassani1999, heckman2005}. In galaxies with extended emission line regions, the total flux of [\ion{O}{3}] may overestimate the current AGN luminosity as measured by other indicators \citep{hainline2013,hainline2014a}. This overestimation is quite important at high redshift, where extended emission is difficult to resolve, especially with current and upcoming near-IR surveys that use [\ion{O}{3}] as an AGN luminosity indicator. By targeting \textit{nearby} AGNs with observed extended emission line regions with extensive multi-wavelength data, we can explore how common indicators of AGN luminosity, such as X-ray and infrared flux, compare to both the nuclear and extended [\ion{O}{3}] luminosity as a function of spectral aperture size. 

In this paper, we focus on a local pair of AGNs, J181611.72+423941.6 and J181609.37+423923.0 (UGC 11185 NED02 and UGC 11185 NED01, and hereafter known as J1816NE and J1816SW) at $z \approx 0.04$ (175 Mpc). Due to the wide separation between these interacting galaxies ($\sim 23$ kpc), observations of these objects provide a relatively rare glimpse of the concurrent growth of merging supermassive black holes at an early merger stage. J1816NE was included in a sample of objects with extended kpc-scale [\ion{O}{3}]$\lambda$5007-emitting clouds in \citet{keel2012}, who used optical spectroscopy for this object to describe the ionization and kinematics for some of the extended gas. Additionally, \citet{keel2015} used narrow- and medium-band Hubble Space Telescope ({\em HST}) imaging and Fabry-Perot spectroscopy for the pair to provide evidence that the extended clouds in J1816NE were oriented in ionization cones potentially excited by the central AGN. Their data did not allow for a full exploration of the ionization as a function of spatial position across these cones, and the determination of this ionization structure is a central focus of this paper. In addition, a \textit{Swift/BAT} source at the position of the J1816 system was detected with a hard X-ray flux $F_{\mathrm{14-195 keV}} = (19\pm4) \times 10^{-12}$ erg cm$^{-2}$ s$^{-1}$ \citep[log($L_{\mathrm{14-195 keV}}/\mathrm{erg\; s}^{-1}) = 43.9$, ][]{baumgartner2013}, and both galaxies were a part of a broad study of the merger and clustering properties of \textit{Swift/BAT} sources in \citet{koss2010}. J1816SW has not been the focus of any detailed analysis, although both galaxies are listed as Seyfert 2 galaxies in the catalogue of \citet{veroncetty2010}. As the two galaxies are only separated by $28\secpoint3$, many of the observations of J1816NE include J1816SW, which provides for a wealth of multi-wavelength data, including {\em Herschel} far-IR data, which we can use to explore the relationship between the AGN and galaxy light to the spectral energy distributions (SEDs) for both galaxies.

In this paper, we use long-slit optical spectroscopy obtained with the OSMOS instrument at MDM observatory of the J1816 merger pair to explore both objects in further detail. The comparison of these two galaxies is vital if we are to understand how AGNs evolve as they go through a major merger. Recent results regarding the variation of AGN luminosity over short timescales \citep[e.g.,][]{hickox2014,schawinski2015} indicate that perhaps galaxies in mergers may cycle between active and inactive states as the merger progresses. The J1816 merger pair offers an ideal test case for exploring this. For the first time, we present an analysis of the ionization properties as a function of spatial position across the face of both galaxies. We also focus on the kinematics of the extended [\ion{O}{3}]-emitting clouds to the southeast and west of the central nucleus in J1816NE to see how the powerful AGN has been driving large kpc-scale ionized high-velocity outflows. While both objects host AGNs, the extended emission line regions seen in J1816NE indicate a much larger intrinsic AGN luminosity than would be predicted by the IR emission for this object, in contrast with what is seen from the compact emission of J1816SW. We further explore the detailed SEDs, X-ray properties, and NLR size as a function of AGN luminosity for both galaxies in this merger pair. 

We describe our observations, data reduction and spectra extraction in Section \ref{sec:observations}, examine the gas ionization structure and kinematics for the system in Section \ref{sec:kinematicsandionization}, and in Section \ref{sec:broadband} we describe our modeling of the SEDs for the objects using a combination of AGN and star-forming galaxy templates. We compare the AGN luminosity indicators for this pair of galaxies in Section \ref{sec:lumindicators}, and we finally conclude in Section \ref{sec:conclusions}. Throughout, we assume a standard $\Lambda$CDM cosmological model with $H_0 = 71$ km s$^{-1}$ Mpc$^{-1}$, $\Omega_{M} = 0.27$, and $\Omega_{\Lambda} = 0.73$ \citep{komatsu2011}.

\section{Observations and Data Reduction}
\label{sec:observations}

\begin{deluxetable}{lccr}
\tabletypesize{\scriptsize}
\tablecaption{Observations \label{tab:observations}}
\tablewidth{0pt}
\tablehead{
\colhead{Observation Target} & \colhead{Date} & \colhead{PA$^{\mathrm{a}}$} & \colhead{Exp. Time} 
}
\startdata

J1816NE & April 19 2013 & 126$^{\circ}$ & 2x1800s \\
J1816NE+SW & April 21 2013 & 59$^{\circ}$ & 1x1800s \\
J1816NE+SW & June 30 2014 & 59$^{\circ}$ & 3x1800s \\
J1816SW & June 30 2014 & 25$^{\circ}$ & 3x1800s

\enddata
\tablenotetext{a}{In degrees east of north.}
\end{deluxetable}     

	\begin{figure}[htp]
	\begin{center}
	\includegraphics[width=0.45\textwidth]{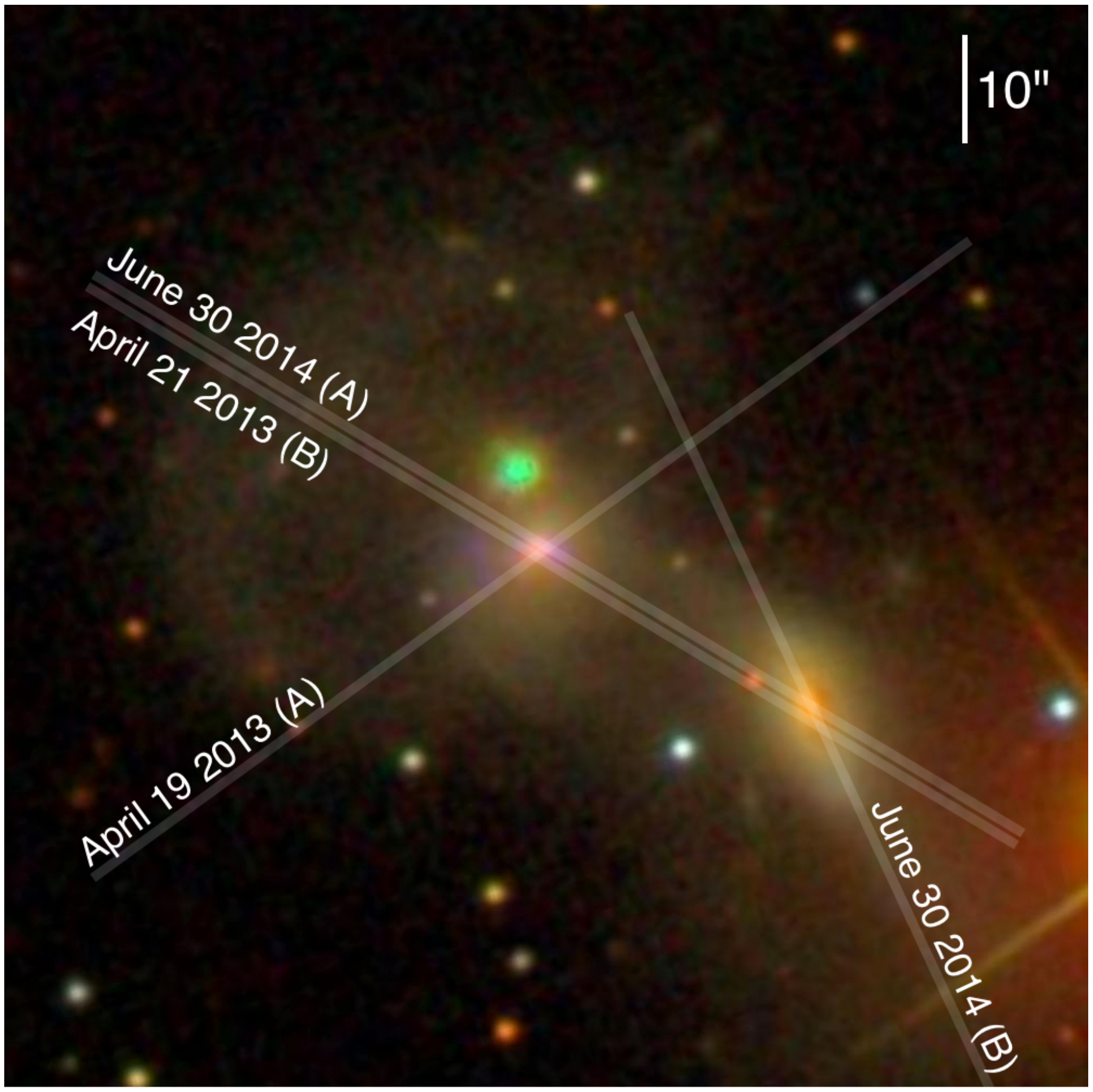}
		\end{center}
	\caption{
	\label{fig:sdssimage} SDSS composite $gri$ image for the J1816 pair. North is up, and east is to the left. J1816NE is found at image center, while J1816SW is located to the southwest. We illustrate the four MDM OSMOS slit positions used in this study with transparent white lines. In J1816NE, as the $gri$ bands have been mapped to blue, green and red \citep{lupton2004}, the haze of purple around the nucleus is strong [\ion{O}{3}] and H$\alpha$+[\ion{N}{2}] emission which can be seen in more detail in Figures \ref{fig:velocity_figure} and \ref{fig:multiwaveimage} .} 
	\epsscale{1.}
         \end{figure}

We observed J1816NE and J1816SW using the OSMOS spectrograph installed on the 2.4m Hiltner Telescope at MDM Observatory over the course of three nights: April 19 and 21 2013, and June 30 2014. We used a total of four slit positions to observe the galaxies, which are shown in Figure \ref{fig:sdssimage}. For each observation, we used a $1\secpoint2$ by 20$'$ slit, with a VPH grism with $R=1600$ (0.7 \AA / pixel). For the April 19 and April 21 observations (2x1800s and 1x1800s, respectively), the wavelength range we used was $3100 - 5960$ \AA, while when we returned to the objects for the June 30 observations (both 3x1800s), we used a wavelength range of $4900 - 6900$ \AA\ to cover the [\ion{N}{2}] and H$\alpha$ emission line region. The seeing was measured to be $1\secpoint6$ for the April 19 observation, $0\secpoint9$ for the April 21 observation, and $1\secpoint5$ for the June 30 observation. Further details of the observations, including position angles, are given in Table \ref{tab:observations}. 

The data were reduced following standard \textit{IRAF}\footnote{\textit{IRAF} is distributed by the National Optical Astronomy Observatory, which is operated by the Associate of Universities for Research in Astronomy (AURA) under cooperative agreement with the National Science Foundation.} routines from the \textit{longslit} package, including bias and dark subtraction, flat fielding, wavelength calibration, and telluric and background subtraction. For flux calibration, we observed the spectrophotometric standards BD+33d2642 and BD+28d4211 from \citet{oke1990} for the April 19/21 and June 30 observations, respectively. 

To calculate the systemic redshift of each of the two galaxies as discussed in the next section, we extracted a spectrum from the central 3\arcsec, where the center of the galaxy was defined based on the peak of the continuum trace between the emission lines. However, in order to understand the ionization properties of the pair as a function of position across each galaxy, we extracted individual one-dimensional spectra along the slit. For both sets of observations, we extracted spectra in bins of 1$\secpoint$6 (6 pixels at $0.273\, ''$/pixel) such that we did not oversample the spectra based on the observed seeing. These spectra were then flux calibrated, extinction corrected, and a heliocentric correction was applied.

\section{Gas Ionization Structure and Kinematics}
\label{sec:kinematicsandionization}

The optical portion of a galaxy's spectrum contains a rich set of emission features which we can use to explore the ionization properties and gas kinematics for this pair of objects. For the slit position aligned from NE to SW (Figure~\ref{fig:sdssimage}), we have observations of both the H$\beta$/[\ion{O}{3}]$\lambda$5007 and H$\alpha$/[\ion{N}{2}]$6583$ emission line regions (from the April 21 and June 30 observations). This allows us to estimate the spatial structure of the ionization state of the gas in both galaxies, which was not possible using the Fabry-Perot maps of the [\ion{O}{3}] line obtained by \citet{keel2015}. Furthermore, for the observations taken during both campaigns, we can use the strong [\ion{O}{3}]$\lambda$5007 emission line to trace the kinematics of gas as a function of position across the interacting system. In order to measure gas kinematics, we require an accurate measurement of the systemic redshift of each of the galaxies. To accomplish this, we used the \textit{GANDALF} \citep{sarzi2006} and \textit{pPXF} \citep{cappellari2004} IDL codes which combine \citet{bc2003} stellar population models and emission lines to the spectra extracted from the central 3'' of each galaxy. From the best-fits, we measured a redshift of $z = 0.0411 \pm 0.0001$ for J1816NE, and $z = 0.0413 \pm 0.0001$ for J1816SW, in agreement with the measurements of \citet{davoust1995}.

	\begin{figure}[htbp]
	\epsscale{1.7} 
	\plotone{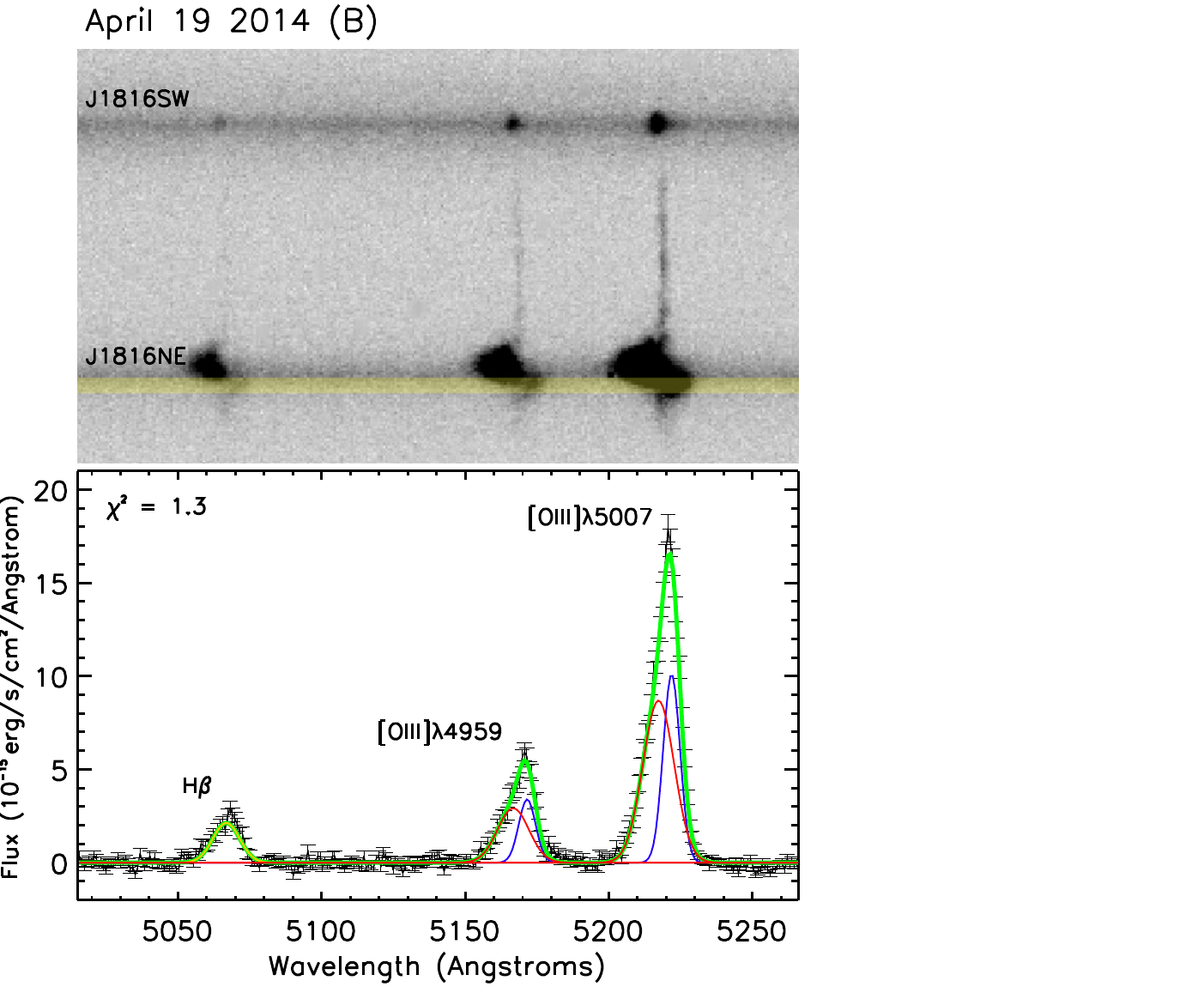}
	\caption{
	\label{fig:twodonedspectra} (Top) Example two-dimensional spectrum from the April 19 2014 (B) observation, as shown in Figure \ref{fig:sdssimage}. The spectra for J1816NE and J1816SW are labelled. In this image, each pixel corresponds to $0\secpoint273$, and the vertical distance between the spectra of the two galaxies is $28\secpoint3$ ($\sim23$ kpc). A clear stream of [\ion{O}{3}]-emitting gas is detected that spans almost the entire distance between the galaxies. (Bottom) One-dimensional extraction from the region shown at top with a yellow bar. In black, we show the observed data, while the total model is plotted in green. We also plot the two fits to the [\ion{O}{3}] emission lines with red and blue lines, and the fit to H$\beta$ with a yellow line.} 
	\epsscale{1.}
         \end{figure}

To understand the gas ionization and kinematics, we fit and removed the continuum in each spectrum, and then modeled the observed emission lines using a series of Gaussians. We fixed the flux ratio of [\ion{O}{3}]$\lambda$5007 to [\ion{O}{3}]$\lambda$4959, as well as [\ion{N}{2}]$\lambda$6583 to [\ion{N}{2}]$\lambda$6548 to the theoretical ratio of 2.98. When the signal-to-noise ratio (S/N) permitted (ie, if two emission line components added to the fit lowered the reduced $\chi^2$ significantly above what was measured with just one component) we used two components to fit each individual emission feature, with the FWHM of the components allowed to vary. Uncertainties on the recovered centroid and flux values were found using a Monte Carlo approach, where we generated 500 artificial spectra by perturbing the flux at each wavelength in the true spectrum by a random amount consistent with the $1\sigma$ error spectrum. We then fit each simulated spectrum using the same procedure as was done on the true spectrum, and the standard deviation of the distribution of centroids and fluxes measured from the artificial spectra was used as the error on the true measurement. In Figure \ref{fig:twodonedspectra}, we show both the two-dimensional and a sample one-dimensional spectrum for the April 21B observation to demonstrate the fitting that was performed.  The [\ion{O}{3}] line profiles across the slit are kinematically complex, and we measured the line centers, the total emission line flux, and the FWHM values from the total model fits to the emission lines. Our FWHM values were deconvolved with the instrumental resolution (for OSMOS, the instrumental resolution is $3.6$\AA) to produce the intrinsic line widths.

\subsection{Gas ionization structure}
\label{sec:ionization}

	\begin{figure}[htbp]
	\epsscale{1.2} 
	\plotone{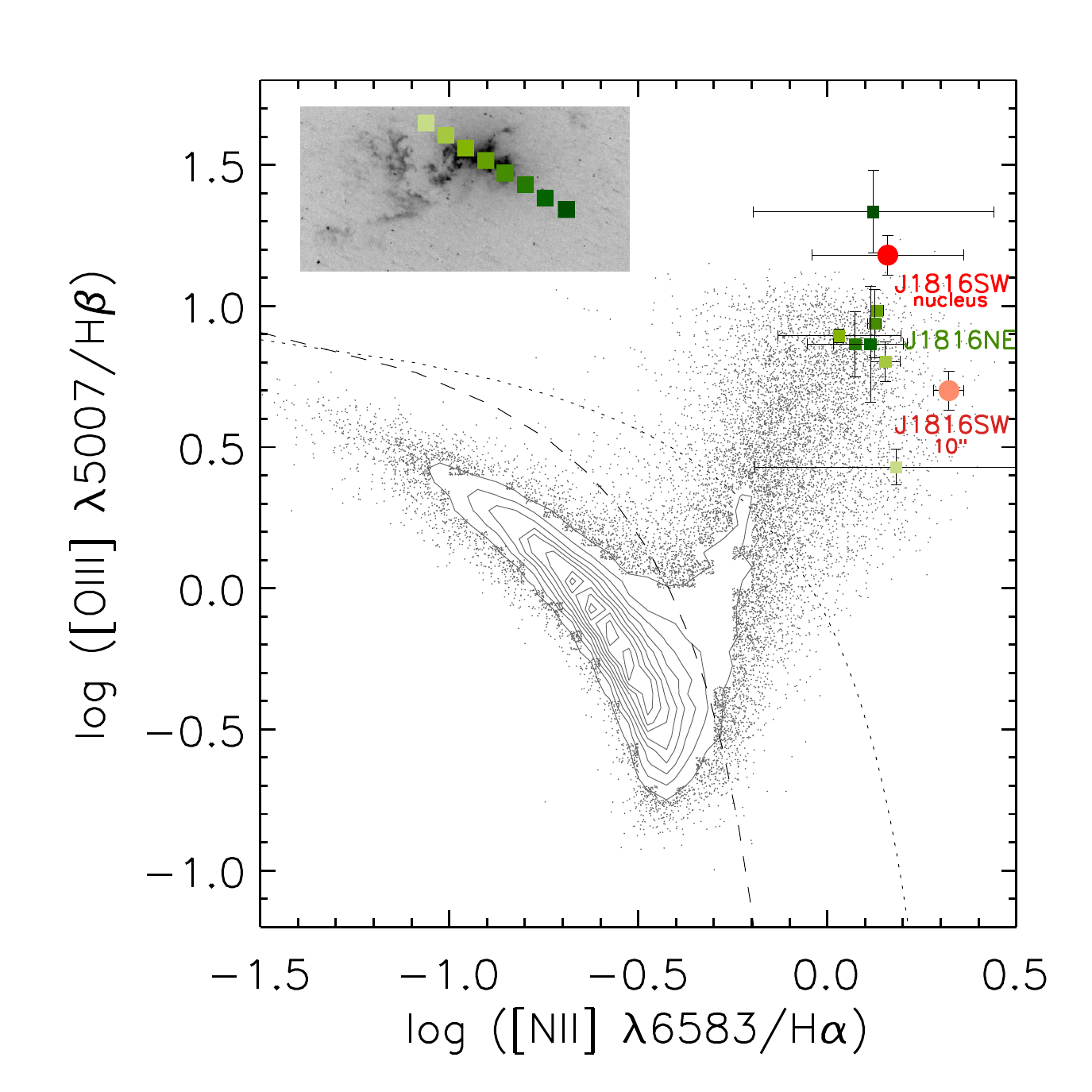}
	\caption{
	\label{fig:bpt} Empirical ``BPT'' diagnostic diagram for J1816NE and B. This diagram uses the strong line ratios of [\ion{O}{3}]$\lambda$5007/H$\beta$ against [\ion{O}{3}]$\lambda$6583/H$\alpha$ to demonstrate gas ionization properties. The gray points and contours represent SDSS local AGN and star-forming galaxies selected from the fourth data release (DR4) of the SDSS \citep{york2000,adelman2006}, and we overplot two curves designed to separate star-forming galaxies and AGN. The dashed line, from \citet{kauffmann2003b}, demarcates star-formation and AGNs empirically, while the dotted line, from \citet{kewley2001a}, represents the limit to the line flux ratios that can be produced by HII regions photoionized by star formation from stellar population synthesis models. In shades of green, we show the positions for the extracted regions across the face of J1816NE (in the inset, we show the positions along the ionization cone where the extractions were made). In red we show the nucleus region of J1816SW. In light-red, we show the ratios for J1816SW using a 10\arcsec\ extraction region on the 2D spectrum, where the resulting position on the diagram better agrees with the underlying comparison sample.} 
	\epsscale{1.}
         \end{figure}

We first present the ionization properties of the galaxies measured across the NE to SW slit position between the two galaxies. We can explore common optical emission line ratios that are used to understand the ionization mechanism of the gas as a function of position across each galaxy. The commonly used ``BPT'' diagnostic diagram \citep{bpt1981} compares the ratio of the strong emission lines [\ion{O}{3}]$\lambda$5007/H$\beta$ against [\ion{N}{2}]$\lambda$6583/H$\alpha$ to separate gas ionized by star formation from gas excited by the presence of an AGN. We show the BPT diagram for J1816NE and J1816SW in Figure \ref{fig:bpt}. Because of the physical extent of the [\ion{O}{3}] emitting regions for J1816NE, we can plot multiple points on the diagram colored as indicated on the inset {\em HST} [\ion{O}{3}] image. The grey points and contours are SDSS local AGN and star-forming galaxies from SDSS DR4 \citep{york2000,adelman2006} for comparison. The dashed line gives the empirical separation between star-forming galaxies (below), and active galaxies (above) from \citet{kauffmann2003b}, while the dotted line shows the limit to the line flux ratios that can be produced by HII regions photoionized by star formation from stellar models from \citet{kewley2001a}. The emission-line ratios for the gas observed along the slit for J1816NE are in a range characteristic of AGNs, and the position of the nucleus on the diagram agrees with what was presented in \citet{keel2012}. We also confirm that the extended kpc-scale emission observed appears to be ionized primarily by the AGN. While further multi-wavelength observations must be taken to obtain a mass outflow rate, these results demonstrate that extent to which powerful AGNs can affect gas throughout their hosts. 

We also plot the position of J1816SW on figure \ref{fig:bpt}. The H$\beta$ line is only detected at $3\sigma$ significance in the central extraction region for J1816SW, and we show the position of J1816SW with a circle. The position of the nuclear extraction of J1816SW has a higher [\ion{O}{3}]$\lambda$5007/H$\beta$ ratio than the SDSS comparison sample, which is expected, as SDSS spectra use 3\arcsec\ diameter fibers that allow for both AGN and galaxy contribution to the emission line fluxes. If we instead use a larger 10\arcsec\ extraction region for our J1816SW spectrum, we calculate flux ratios that better agree with the SDSS objects, and we show this position in light red on Figure \ref{fig:bpt}. Based on the observed line ratios, J1816SW is also an optical AGN, but with a much more compact NLR. 

\subsection{Gas kinematics}
\label{sec:kinematics}

	\begin{figure*}[htbp]
	\centering
	  \begin{tabular}{@{}c@{}}
	   \includegraphics[width=.95\textwidth]{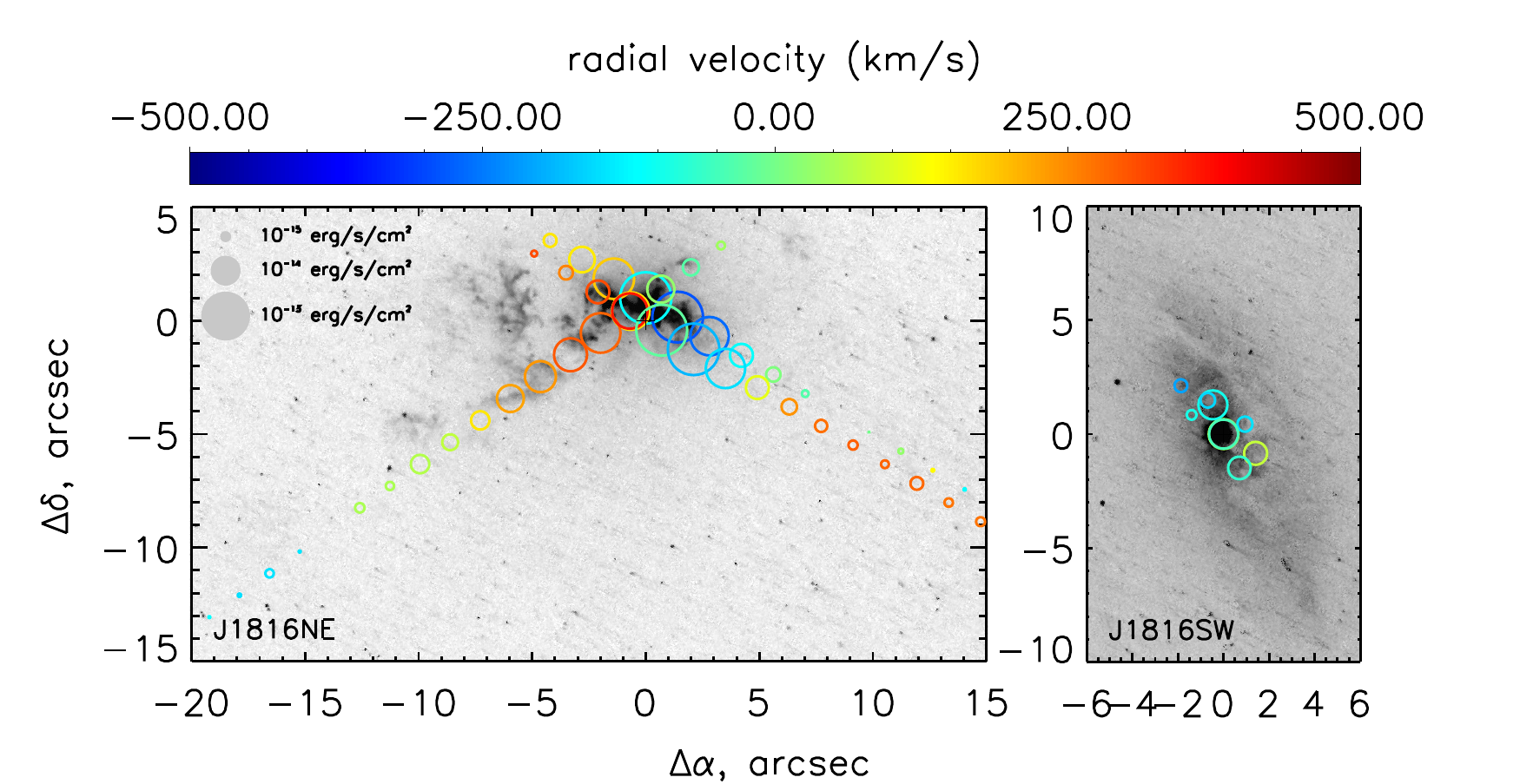} \\
	   \includegraphics[width=.95\textwidth]{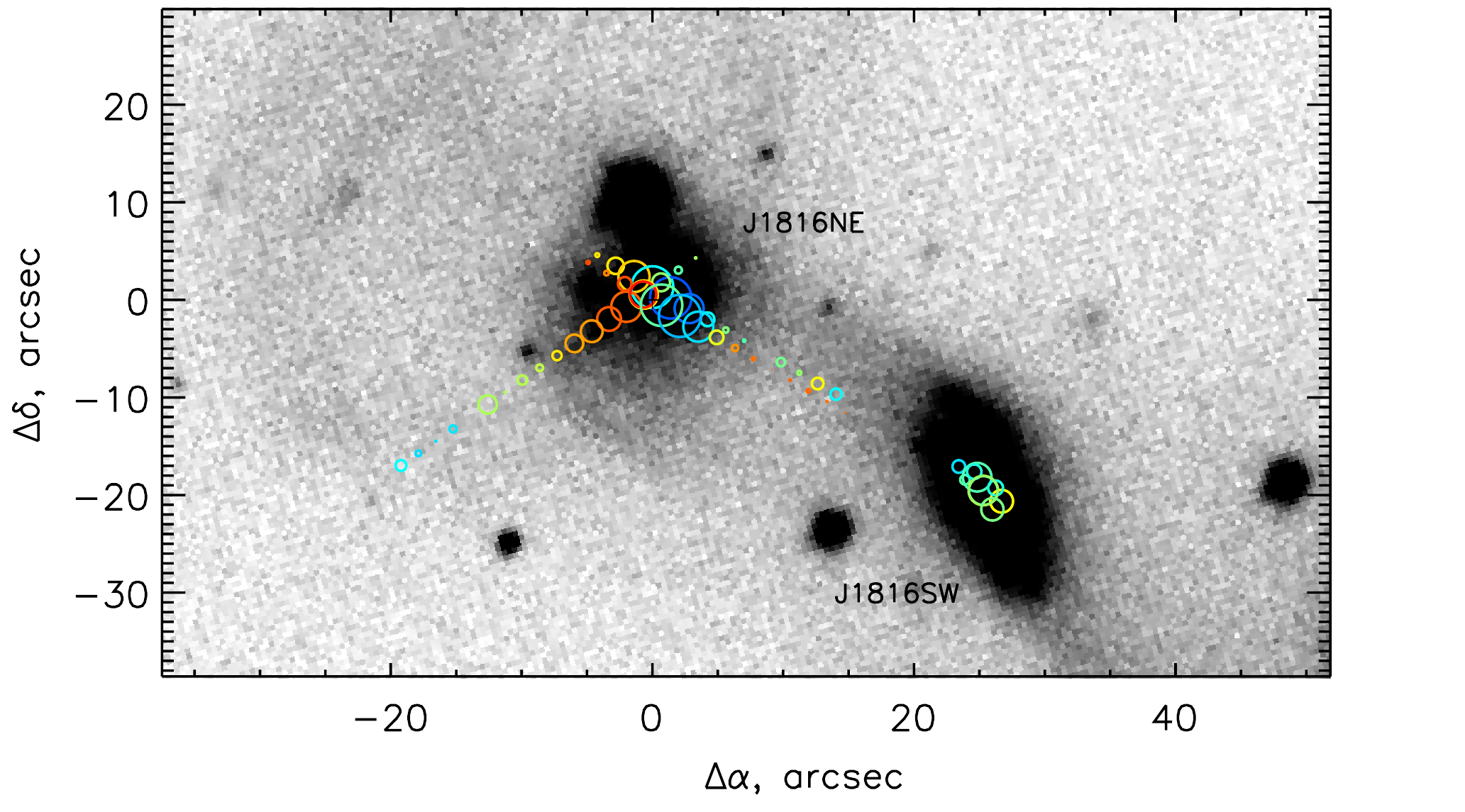} \\
	\end{tabular}
	\caption{
	\label{fig:velocity_figure} (Top) [\ion{O}{3}]$\lambda$5007 radial velocity measured for three slit positions across the face of J1816NE (left) and J1816SW (right) with north up and east to the left. We show the {\em HST} ramp-filter images focused on the portion of the spectrum containing the [\ion{O}{3}] emission line in greyscale. Circles overlaid on each image represent measurements at each of our extracted positions along each slit shown in Figure \ref{fig:sdssimage}. The size of the circle is proportional to the flux in the [\ion{O}{3}]$\lambda$5007 emission feature as shown in the key in the upper left, while the circles are colored according to the velocity of the centroid with respect to the systemic velocity of each galaxy, as shown in the color bar at the top of the figure. The measured velocities indicate that we are observing the NLR bicone in J1816NE, with the side opening towards the Earth on the west side. We measure strong [\ion{O}{3}]$\lambda$5007 emission at near systemic velocity along the entirety of the stream between J1816NE and J1816SW. The J1816SW NLR is more compact, and has less velocity structure, with blueshifted [\ion{O}{3}] as compared to the systemic velocity. (Bottom) Similar to the top panel, but here we overplot the kinematic results on the SDSS $g$ band image of both galaxies to compare to the faint gas extending between J1816NE and J1816SW.} 
        \end{figure*}

\begin{figure*}[tbp]
	\epsscale{1.0}
	\plotone{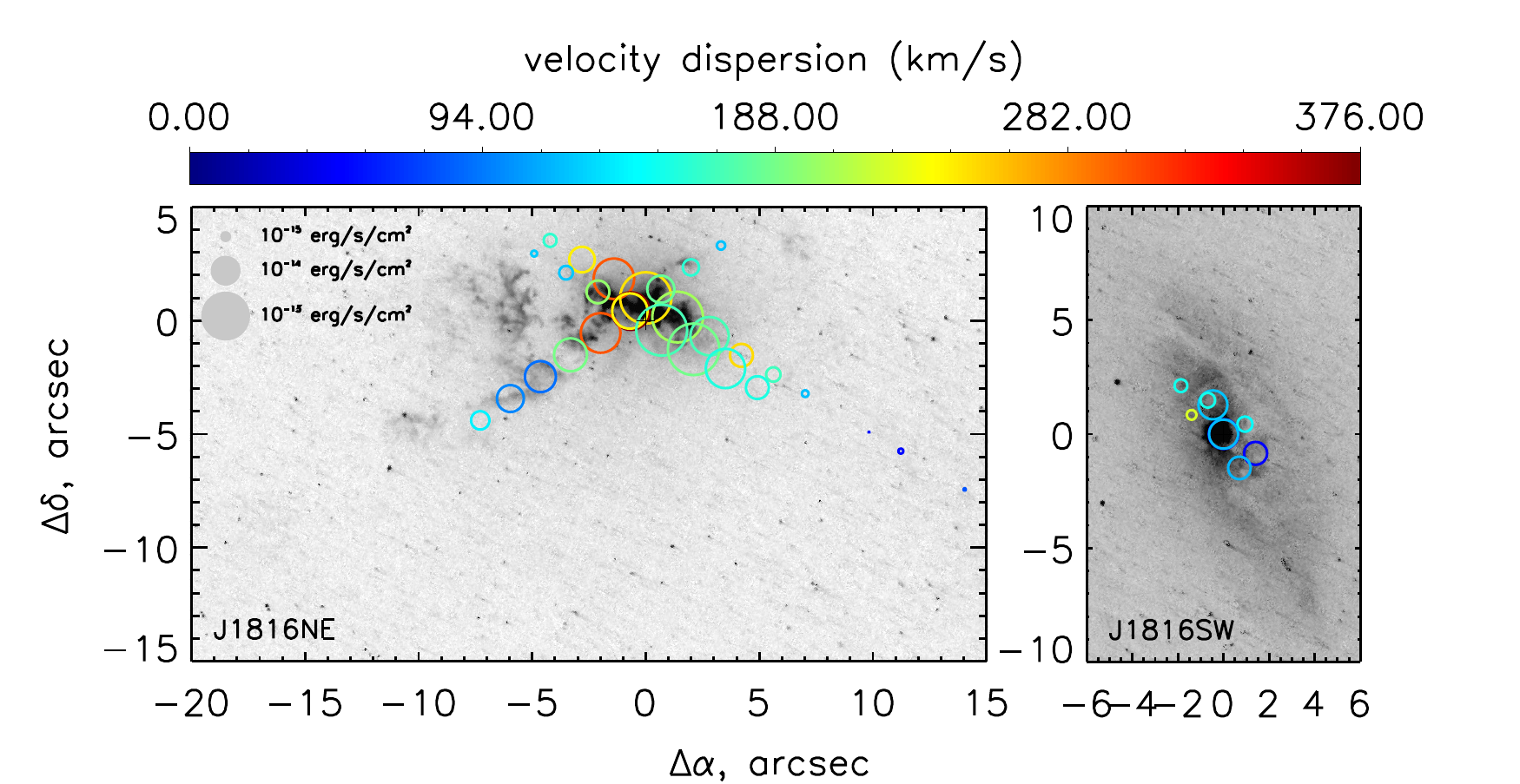}
	\caption{
	\label{fig:sigma_figure} [\ion{O}{3}]$\lambda$5007 velocity dispersions measured for three slit positions across the face of J1816NE (left) and J1816SW (right), with north up and east to the left. We show the {\em HST} ramp-filter image focused on the portion of the spectrum containing the [\ion{O}{3}] emission line in grey. Circles overlaid on each image represent measurements at each of our extracted positions along each slit shown in Figure \ref{fig:sdssimage}. The size of the circle is proportional to the flux in the [\ion{O}{3}]$\lambda$5007 emission feature as shown in the key in the upper left, while the circles are colored according to the velocity dispersion, as shown in the color bar at the top of the figure.} 
	\epsscale{1.}
         \end{figure*}

We next use observations from both MDM campaigns and all slit positions to examine the gas kinematics across the system, obtained from the velocities observed in the [\ion{O}{3}] line. In the top panel of Figure \ref{fig:velocity_figure} we overplot the observed velocities on top of the {\em HST} ramp-filter [\ion{O}{3}] image of J1816NE\footnote{The HST images shown in Figures \ref{fig:velocity_figure}, \ref{fig:sigma_figure}, \ref{fig:bpt}, and \ref{fig:multiwaveimage}, and used in the analysis in Section \ref{sec:lumindicators} are taken from HST Proposal 12525 (Keel et al.). The continuum image we use is from the WFC3 UVIS2 instrument, taken through the F621M filter, while the [\ion{O}{3}] image is from the ACS WFC1 instrument with a FR505N ramp filter.}. For this figure, the radius of each circle corresponds to the relative (logarithmic) flux in the emission line, while the circles are colored according to the velocity, as shown with the color-bar above the figure. This figure demonstrates that within a radius of $\sim5''$ centered on J1816NE, there exists a strong radial velocity gradient across the galaxy such that gas to the west is blueshifted at $\sim 300$ km s$^{-1}$, while the gas to the east is redshifted at $\sim 300$ km s$^{-1}$. We interpret these velocities as demonstrating that the AGN ionization cone observed in the {\em HST} narrow-band imaging is opened towards our line of sight to the west, and opened away from our line of sight to the east. For the April 19A spectrum, our observations demonstrate that while the gas close to the nucleus is redshifted, along the slit to the southeast the gas slows and moves back towards the systemic velocity, even out to $> 15$ kpc from the nucleus. The BPT line diagnostics described above suggest that the full extent of this line emission is powered by photoionization from an AGN, showing that, in addition to producing an outflowing cone of highly ionized gas, the AGN in J1816NE can influence the state of the gas {\em between} the galaxies that is extended along a tidal bridge induced by the merger. In the bottom panel, we show these kinematic results overplotted on the SDSS $g$ band image for comparison with the faint gas surrounding the merging pair of galaxies. 

In the April 21B spectra, we further observe that this stream of ionized gas spans almost the entire distance between J1816NE and J1816SW (as seen in the top panel of Figure \ref{fig:twodonedspectra}). In Figure \ref{fig:velocity_figure}, it can be seen that this gas is slightly redshifted from the systemic redshift, and, based on the redshift of J1816SW, we are observing the gas that bridges the two merging galaxies (which is kinematically distinct from the outflow in the ionization cones) become ionized during the interaction. These transitions in the velocity to either side of the nucleus indicate a kinematic size of the NLR of around $5''$ (4 kpc), in agreement with what we measure from the [\ion{O}{3}] surface brightness in Section \ref{sec:lumindicators}. 

In Figure \ref{fig:sigma_figure}, we also plot the [\ion{O}{3}] velocity dispersion across the face of J1816NE, again with the size of each point corresponding to the relative flux in the emission line and the points colored according to the velocity given in the color bar at the top of the figure. The western bicone opening shows velocity dispersions of $\sim 200$ km s$^{-1}$, which increases to $> 300$ km s$^{-1}$ near the galaxy center. These velocities and velocity dispersions agree with the results for J1816NE from \citet{keel2015}, who provided full BTA Fabry-Perot maps for [\ion{O}{3}]$\lambda$5007 for the pair. These results support the growing evidence for the existence of powerful, fast-moving outflows in Type 2 AGNs across a wide range in AGN luminosities. \citet{liu2013}, \citet{harrison2014} and \citet{mcelroy2015} all demonstrated evidence for fast-moving ionized outflows in their samples of Type 2 AGNs at larger [\ion{O}{3}] luminosities. Based on the merger state of the J1816 system, it is possible that as more gas is driven towards the nucleus of J1816NE, the accretion rate onto the black hole and the bolometric luminosity will increase, leading to larger future outflow velocities. 

Using a similar methodology, we also measured the [\ion{O}{3}] kinematics for J1816SW, which we show in the right panels in Figures \ref{fig:velocity_figure} and \ref{fig:sigma_figure}. J1816SW has a very compact emission line region, and we only detected [\ion{O}{3}] in the central three extraction bins (a region spanning 4$\secpoint$8, or 3.9 kpc), with the emission line being significantly stronger in the central 1$\secpoint$6 (1.3 kpc) region. Unlike the kinematics observed for J1816NE across the NLR, the velocity observed for the nucleus of J1816SW indicates an overall blueshift on the order of $100 - 200$ km s$^{-1}$ with respect to the systemic velocity of the system, with velocity dispersions of $\sim100$ km s$^{-1}$. For such a compact NLR in J1816SW, if a biconic outflow exists similar to what is observed in J1816NE, we are only observing the blueshifted gas on one side.

\section{Broad-band Multi-wavelength Analysis}
\label{sec:broadband}

	\begin{figure*}[htbp]
	\epsscale{1.0} 
	\plotone{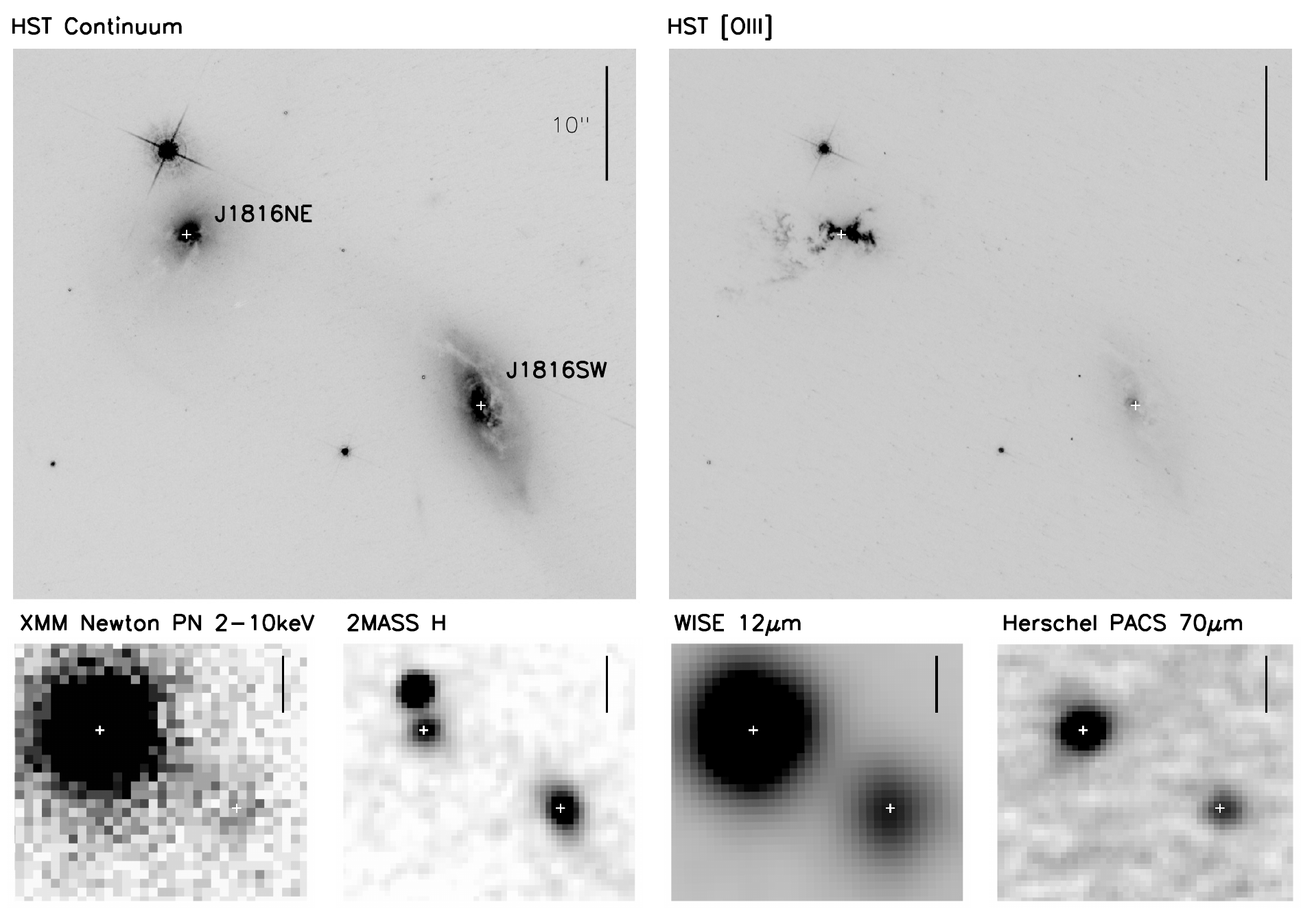}
	\caption{
	\label{fig:multiwaveimage} Multi-wavelength images of J1816NE and J1816SW. The top panels show the optical continuum image ({\em HST} WFC3 UVIS, F621M filter, left) and the [\ion{O}{3}]$\lambda$5007 ramp-filter image (ACS WFC1, FR505N filter, right). Below, we show the \textit{XMM-Newton} PN $2-10$ keV (far left), 2MASS $H$-band (left), WISE ``W3'' (12$\mu$m, right), and {\em Herschel} PACS 70$\mu$m images (far right) for both objects. In all images, north is up, and east is to the left, and crosses mark the centers of J1816NE and J1816SW in each image. We also show a 10\arcsec\ scale bar in each panel. The sizes of the [\ion{O}{3}]-emitting regions for J1816NE and J1816SW are vastly different, as shown in the top two panels.} 
	\epsscale{1.}
         \end{figure*}

Because of the proximity of the J1816 pair of galaxies, the pair has been targeted with observations across a wide wavelength range, as shown by the optical, X-ray, and infrared images shown in Figure \ref{fig:multiwaveimage}. In Section \ref{sec:SEDs}, we introduce these observations and discuss the results from fitting multiple templates to the spectral energy distributions for each object, and in Section \ref{sec:Xray} we describe the analysis of the X-ray observations. In Section \ref{sec:lumindicators}, we compare luminosity indicators derived from these models, as well as the observed spectra. 

\subsection{Modeling Spectral Energy Distributions}
\label{sec:SEDs}

Both J1816NE and J1816SW have existing optical through infrared photometry derived from various sources. We used SDSS ``model'' magnitudes for the optical \textit{ugriz} photometry from the SDSS Data Release 9 \citep{ahn2012}. For the near-IR photometry, we used 2MASS \textit{JHK} images, and then used SExtractor \citep{bertin1996} to estimate Kron-like elliptical aperture magnitudes for both objects (``MAG\_AUTO''), as these were extended sources. In the mid-infrared we used WISE photometry, which was taken in four bands: W1 (3.4 $\mu$m), W2 (4.6 $\mu$m), W3 (12 $\mu$m), and W4 (22 $\mu$m). Emission from dust heated by the accretion disk in an AGN results in a characteristic red infrared slope that is well-sampled with WISE photometry. For these data, we followed the recommendation of the Explanatory Supplement to the AllWISE Data Release Products\footnote{http://wise2.ipac.caltech.edu/docs/release/allwise/expsup/index.html} and used the $w*mag$ photometry for J1816NE and $w*gmag$ photometry\footnote{We use the $w*gmag$ magnitudes for J1816SW as this object is a part of the 2MASS Extended Source Catalog, and these magnitudes use elliptical apertures to better capture the total source brightness. The $w*mag$ and $w*gmag$ photometry differs on average by only 0.2 mag.} for J1816SW taken from the AllWISE Source Catalog (Cutri et al. 2013). 

Far-IR data is helpful for constraining the AGN and star-formation contribution to each objects SED. J1816NE, by virtue of being a \textit{Swift/BAT}-detected source, was targeted by the {\em Herschel} Space Observatory in the far-infrared with the {\em Herschel} Photoconductor Array Camera and Spectrometer (PACS) instrument at 70 and 160 $\mu$m ({\em Herschel} Program OT1\_rmushot\_1, PI: R. Mushotzky). Fluxes at these wavebands were estimated for J1816NE in \citet{melendez2014}, and for J1816SW (which was observed in the same field due to its proximity to J1816NE) we used the HIPE software to extract aperture fluxes for this object. (We verified our flux measurement method by also extracting the flux for J1816NE, which agreed with the value of \citealt{melendez2014}.) Following \citet{balog2014}, these fluxes were corrected for extraction aperture flux losses, and uncertainties were obtained using the standard deviation of the background measured in apertures around the source, with a 5\% photometric uncertainty added in quadrature (following the PACS Observer's Manual\footnote{http://herschel.esac.esa.int/Docs/PACS/html/pacs\_om.html}). 

	\begin{figure}[htbp]
	\epsscale{1.2} 
	\plotone{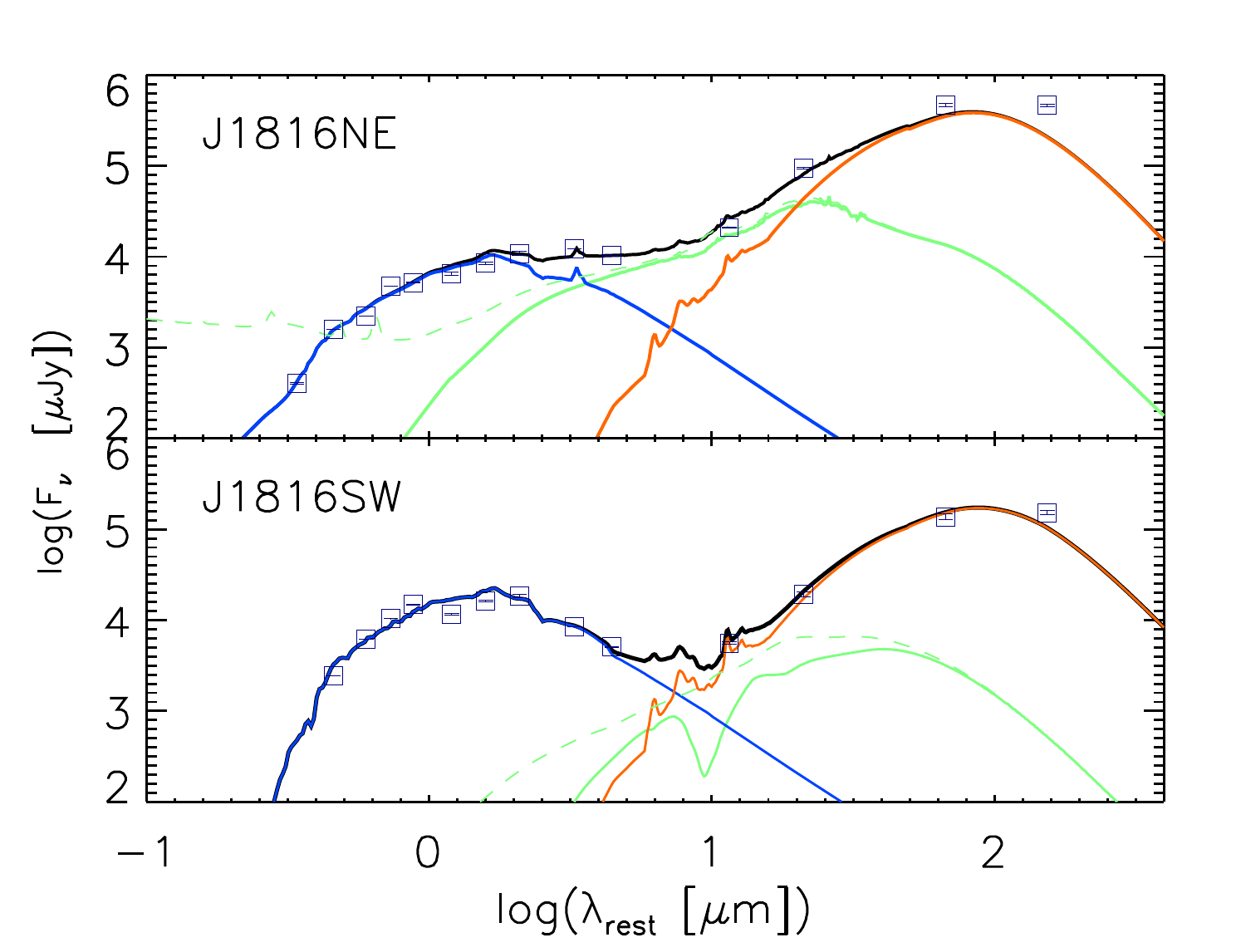}
	\caption{
	\label{fig:sed_results} SED modeling results for J1816NE (top) and J1816SW (bottom). These objects were modeled with a combination of three sets of templates, a galaxy template (blue), an AGN template (green), and an IR component from cold dust heated by star formation (red). We also plot, with a dashed green curve, the AGN template with the effects of dust obscuration removed. The data points are shown with squares, with error bars given for the flux. Due to the central obscuration, both galaxies only show a significant contribution from AGN emission at around 10 $\mu$m.} 
	\epsscale{1.5}
         \end{figure}

Using these optical through far-IR data, we fit galaxy, AGN, and starburst templates to the broad-band SEDs for both J1816NE and B. To model the stellar continuum contribution to these objects, we used the empirically-derived galaxy models from \citet{assef2010}, and summed the elliptical, Sbc, and irregular galaxy templates with individual coefficients to produce the final template, following \citet{chen2015}. For the AGN contribution, we used the \citet{netzer2007a} and \citet{mullaney2011} AGN templates, where we have applied foreground extinction to simulate the effects of dust obscuration from the torus in these objects. To model the far-IR photometry we have also used the \citet{charyelbaz2001} starburst templates, which account for cold dust heated by galaxy-scale star formation. We used a $\chi^2$ minimization algorithm to find the best-fitting set of coefficients and AGN extinction to fit the observed SEDs. We provide the luminosities ($\nu L_\nu$) calculated at 1$\mu$m in the rest frame for the best-fitting galaxy and AGN templates in Table \ref{tab:sedresults} and we plot the SEDs in Figure \ref{fig:sed_results}. In this Figure, we can see that the optical portion of each galaxy's SED is dominated by the \citet{assef2010} Sbc (for J1816NE) and elliptical (for J1816SW) templates, with AGN emission only becoming strong in the near-IR, and with emission from cold dust dominating the the mid- to far-IR. For J1816NE, the best-fitting AGN template was the \citet{netzer2007a} template, while for J1816SW, the best-fitting AGN template was from \citet{mullaney2011}. From this modeling, J1816NE hosts a significantly (30 times) more luminous AGN than J1816SW, with $L_{\mathrm{MIR}} = (1.9 \pm 0.2) \times 10^{43}$ erg s$^{-1}$ for J1816NE and $(2.9 \pm 0.3) \times 10^{42}$ erg s$^{-1}$ for J1816SW, where $L_{\mathrm{MIR}}$ is calculated as the average luminosity ($\nu L_\nu$) between rest-frame 12-15$\mu$m from the best-fitting AGN template (after removal of reddening), with uncertainties estimated from the fractional uncertainties on the WISE fluxes. From the E(B-V)$_{\mathrm{AGN}}$ values, J1816SW is more dust obscured than J1816NE. The dust could arise from structures in the nucleus or on larger scales in the host galaxy \citep[e.g.][]{goulding2012}; the {\em HST} continuum image for J1816SW has evidence for a strong dust band across the center of J1816SW. We integrated the 8--1000 $\mu$m luminosity for both objects, as derived from the \citet{charyelbaz2001} templates, and measured $L_{\mathrm{FIR}} = (2.5 \pm 0.3)\times 10^{10}$ $L_\sun$ for J1816NE, in agreement with \citet{keel2012}, and $L_{\mathrm{FIR}} = (1.0 \pm 0.3)\times 10^{10}$ $L_\sun$ for J1816SW (our uncertainties were estimated from the uncertainties on the measured {\em Herschel} fluxes). We noted that the IR luminosities we measure are not high enough for either J1816NE or J1816SW to be considered Luminous Infrared Galaxies (LIRGs), which is likely due to the dust in the system not being warm enough to produce infrared emission above the value of $10^{11}$ $L_\sun$ used to define these objects \citep{sandersmirabel1996}. If we assume that this FIR luminosity is entirely due to the heating of dust by star formation, we can use the \citet{kennicutt1998} conversion to calculate the SFR for each object. We estimate 4 M$_{\sun}$ yr$^{-1}$ for J1816NE and 2 M$_{\sun}$ yr$^{-1}$ for J1816SW. The J1816NE fit underpredicts the far-IR flux, although, if we use the conversion between 70$\mu$m PACS flux to $L_{\mathrm{FIR}}$ from \citep{galametz2013} (assuming the entirety of the flux is due to star formation), we calculate $L_{\mathrm{FIR}} = 3.8 \times10^{10}$ $L_\sun$, which corresponds to a SFR of 7 M$_{\sun}$ yr$^{-1}$. If we repeat this calculation for J1816, we obtain $L_{\mathrm{FIR}} = 1.1 \times10^{10}$ $L_\sun$, which corresponds to a SFR of 2 M$_{\sun}$ yr$^{-1}$, in agreement with the far-infrared luminosity and SFR estimated from the SED fitting for this object. 

\begin{deluxetable*}{lcccccc}
\tabletypesize{\scriptsize}
\tablecaption{SED Modeling Results \label{tab:sedresults}}
\tablewidth{0pt}
\tablehead{
\colhead{Object} & \colhead{Sbc$^{\mathrm{a}}$} & \colhead{E$^{\mathrm{a}}$} & \colhead{Im$^{\mathrm{a}}$} & \colhead{AGN$^{\mathrm{b}}$} & \colhead{$E(B-V)_{\mathrm{AGN}}$} & \colhead{Starburst$^{\mathrm{c}}$}
}
\startdata
J1816NE & 43.85 & 42.41 & - & 43.21 & 1.59 & $2.5\times 10^{10}$ $L_\sun$ \\
J1816SW & - & 44.23 & - & 41.72 & 9.99 & $1.0\times 10^{10}$ $L_\sun$
\enddata
\tablenotetext{a}{$\log(\nu L_\nu / \mathrm{erg\, s^{-1}})$ at 1$\mu$m for the \citet{assef2010} Sbc, E, and Im templates.}
\tablenotetext{b}{$\log(\nu L_\nu / \mathrm{erg\, s^{-1}})$ at 1$\mu$m for the \citet{netzer2007a} (for J1816NE) and \citet{mullaney2011} (for J1816SW) AGN templates.}
\tablenotetext{c}{Total Far-IR luminosity derived from fitting the \citet{charyelbaz2001} starburst templates. For J1816NE, we used the \citeauthor{charyelbaz2001} template with index 91, and for J1816, we used the template with index 76.}
\end{deluxetable*}

\subsection{X-ray analysis}
\label{sec:Xray}

In addition to the optical through far-IR data, we have X-ray fluxes from both \textit{Swift/BAT} and \textit{XMM-Newton} observations of both targets. The 14--195 keV flux, as taken from the \textit{Swift/BAT} 70-Month Hard X-ray Survey catalogue \citep{baumgartner2013}, is $(19\pm4) \times 10^{-12}$ erg cm$^{-2}$ s$^{-1}$. The width of the BAT point spread function is $22\minpoint5$, and, as J1816NE and J1816SW are separated by $28\secpoint3$ ($\sim 23$ kpc), the  \textit{Swift/BAT} flux is most likely a combination of the X-ray flux from both objects. 

\begin{figure}[htbp]
	\begin{center}
	\includegraphics[width=0.45\textwidth]{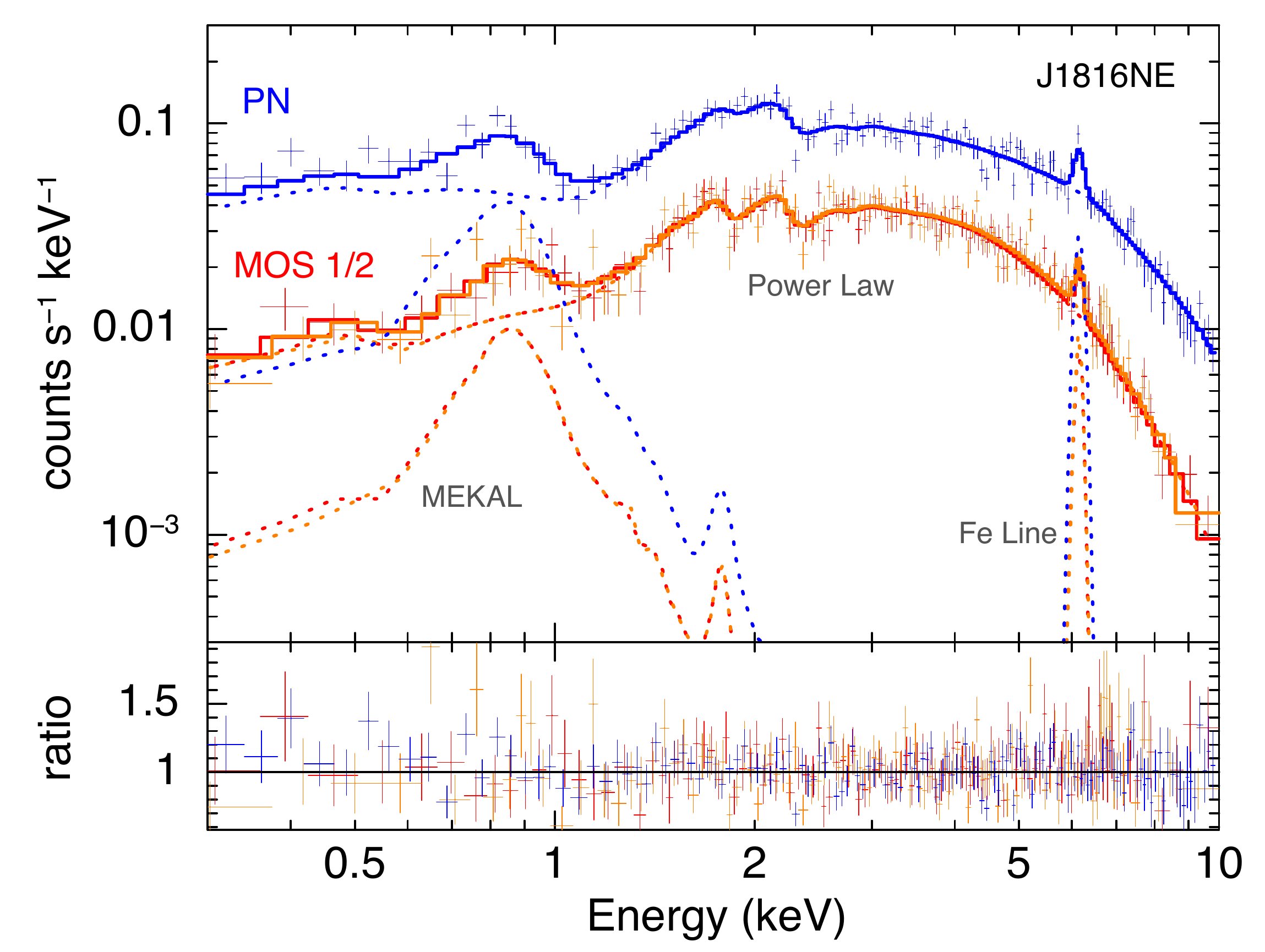}
		\end{center}
	\caption{
	\label{fig:xray_fig} {\em XMM-Newton} X-ray spectra, best-fit model, and residuals for J1816NE. Spectra are shown at observed-frame energies for the pn (black), MOS1 (red) and MOS2 (orange) detectors. The best-fit model includes a power law with partial covering absorption, a thin-thermal (MEKAL) component, and an unresolved Gaussian feature corresponding to the Fe K$\alpha$ line. The model provides an excellent fit to the data and allows an estimate of the intrinsic X-ray luminosity of the source. } 
	\epsscale{1.5}
         \end{figure}

\textit{XMM-Newton}, which has a much higher spatial resolution ($\sim6''$ PSF), can resolve the separate emission from J1816NE and B. These objects were observed as part of two campaigns targeting objects with extended NLRs (PI: Schawinski, OBSID 0672660401 and 0672660501). For our analysis, we only used those observations from the campaign with the longer exposures (OBSID 0672660401) where the object was observed for $\sim 26$ ksec. After cleaning for background flares and applying standard event quality cuts (yielding an effective exposure of 16.5 ks), we extracted a spectrum in the 0.3--10 keV band for J1816NE from each of the MOS1, MOS2, and pn detectors on {\em XMM}. We used extraction regions centered on the source of radius 23\arcsec\ and a background source-free regions adjacent the source of radius 117\arcsec. The spectra are shown in Figure \ref{fig:xray_fig}, and contain 4220, 4394, and 9372 0.3--10 keV counts in the MOS1, MOS2, and pn respectively. For all detectors, background was $<$2\% of the total source counts. We also extracted response (RMF and ARF) files, including a correction for the energy encircled fraction ($\textrm{EEF} \approx 80$\% for the 23\arcsec\ extraction region) in the calculation of ARF. 

J1816NE exhibits a hard X-ray spectrum at energies $<10$ keV; a simple estimate using the hardness ratio, defined as $\mathrm{HR} = (F_{2-10\mathrm{keV}} - F_{0.5-2\mathrm{keV}})/({F_{2-10\mathrm{keV}} + F_{0.5-2\mathrm{keV}}})$, yields $HR = 0.554 \pm 0.008$ for J1816NE, and a simple unabsorbed power-law fit to the 1--10 keV spectrum (while a very poor fit, with reduced $\chi^2 \approx 7$) returns an X-ray photon index of $\Gamma \sim 0.1$, far harder than the  unabsorbed AGN power laws for typical AGN, which lie in the range $1.4<\Gamma<2.2$ \citep[e.g.,][]{tozzi2006}. Fitting the full 0.3--10 keV spectrum, we find that it is well-described by a simple model consisting of a power-law component with partial covering absorption to model the direct nuclear emission from the AGN, an unresolved Gaussian at $\sim6.4$ keV (rest-frame) to model the Fe K$\alpha$ emission line that is common in AGN \citep[e.g.,][]{ricc14fek}, and thin-thermal (MEKAL; \citealt{mewe85, liedahl95}) component to account for soft emission from star formation processes in the host galaxy \citep[e.g.,][]{mine12xraysf}. The total model is modified by Galactic absorption with $N_{\rm H} = 3\times10^{20}$ cm$^{-2}$ \citep{kalb05nhgal}. This fit yields a reduced $\chi^2 = 0.98$ for 363 degrees of freedom. For the power law component the best-fit parameters are photon index $\Gamma = 1.43\pm0.03$, $N_{\rm H} = (2.96\pm0.09)\times10^{22}$ cm$^{-2}$, and covering fraction $f_{\rm cov} = 0.963\pm0.004$, with an unabsorbed rest-frame 2--10 keV power-law flux of $F_{\rm 2-10\; keV} = (5.93\pm0.07)\times10^{-12}$ erg cm$^{-2}$ s$^{-1}$. For the MEKAL component, $kT = 0.69\pm0.04$ keV, with rest-frame 0.5--2 keV flux of $F_{\rm MEKAL} = (2.8\pm0.4)\times10^{-14}$ erg cm$^{-2}$ s$^{-1}$, while the Fe line has $E_{\rm Fe} = 6.41\pm0.02$ keV and flux $F_{\rm Fe} = (1.05\pm0.16)\times10^{-13}$ erg cm$^{-2}$ s$^{-1}$. Assuming a Milky Way gas to dust ratio of $N_H / A_V = 2 \times 10^{21}$ cm$^{-2}$, we obtain $A_V \sim 15$, of similar order to that determined for the AGN component from the SED fitting.

We stress that while this simple phenomenological model likely does not represent a complete physical description of the X-ray emission from this source, it gives a reliable estimate of the luminosity of the hard power-law component which is the primary focus of this work. If we restrict our fits to energies $>$1.5 keV, thus avoiding the MEKAL and unabsorbed soft emission, and apply a simple absorbed power law with the canonical $\Gamma=1.8$ and the Fe emission line, the unabsorbed flux agrees to within $<$1\%. Finally, we note that an extrapolation of our best-fit model to high energies returns a 14--195 keV flux of $15\times10^{12}$ erg cm$^{-2}$ s$^{-1}$, consistent with the \textit{Swift/BAT} measurement \citep{baumgartner2013}.

J1816SW did not yield sufficient source counts to extract a high S/N spectrum, so we restricted our estimates of the flux from the net observed counts. We extracted source counts from a region of radius 8\arcsec\ centered on the object, and background counts from an annulus around the position of J1816NE with radius of 26\arcsec\ (equal to the separation between J1816NE and B), to account for any scattered light from J1816NE. For this fainter source we focus primarily on the pn detector and do not perform extensive light curve cleaning for flares in order to maximize the number of counts. We obtained $149\pm17$ and $25\pm8$ net counts in the 2--10 keV and 0.5--2 keV bands, respectively, yielding $HR = 0.71 \pm 0.09$, implying a hard spectrum similar to that of J1816NE. Assuming the same counts to flux conversion obtained for J1816NE, and accounting for the energy-encircled fraction of 45\% for the 8\arcsec\ extraction region, yields an intrinsic 2--10 keV flux of $F_{\rm 2-10\; keV} = (1.7\pm0.2)\times10^{-13}$ erg cm$^{-2}$ s$^{-1}$. Finally, we note that the relatively small flux of J1816SW implies that any flux scattered into the source aperture of J1816NE for the spectral analysis is negligible ($\approx0.5\%$ in the 2--10 keV band).

\subsection{A Comparison of AGN Luminosity Indicators}
\label{sec:lumindicators}

	\begin{figure*}[htbp]
	\epsscale{1.0} 
	\plotone{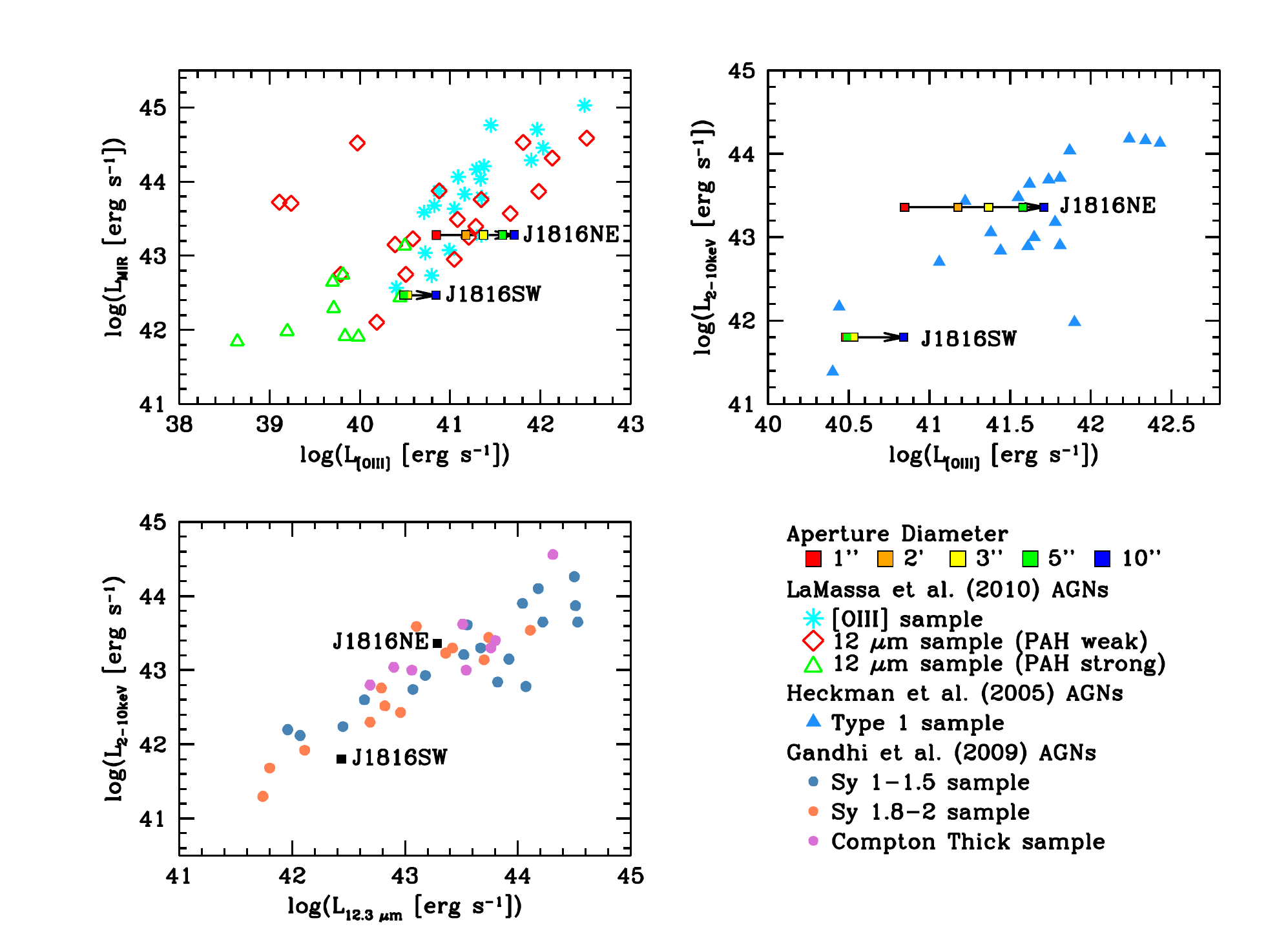}
	\caption{
	\label{fig:oiii_vs_ir} A comparison of luminosity indicators for J1816NE and J1816SW. We compare the luminosities for our objects with those from the literature plotting Type 1 and Type 2 AGNs from \citet{lamassa2010,lamassa2011,heckman2005} and \citet{gandhi2009} with points labelled in the bottom-right. In the top-left panel we plot the [\ion{O}{3}] luminosity against mid-IR luminosity, and in the top-right panel we plot the unabsorbed 2-10 keV X-ray luminosity against [\ion{O}{3}] luminosity. For these two panels, J1816NE and J1816SW are shown with colored squares, where the color indicates the diameter of the virtual circular aperture that was used to estimate the [\ion{O}{3}] flux, 1'' (red), 2'' (orange), 3'' (yellow), 5'' (green), and 10'' (blue). In the bottom-left panel, we plot the unabsorbed 2-10 keV X-ray luminosity against the monochromatic 12.3$\mu$m luminosity. J1816NE, which contains the more luminous AGN, has extended [\ion{O}{3}] emission which changes the total [\ion{O}{3}] luminosity by an order of magnitude as the size of the virtual aperture used to extract the flux increases. The extended emission pushes this object away from the central locus of local AGNs in the [\ion{O}{3}] luminosity vs. mid-IR luminosity relation, and \emph{towards} the central locus of local AGNs in the [\ion{O}{3}] luminosity vs. 2-10 keV X-ray luminosity relation. The X-ray and mid-IR luminosities agree with the comparison samples. J1816SW, on the other hand, has a relatively compact NLR, and the object's position is along the primary locus of local AGNs in all panels.} 
        \epsscale{1.5}
         \end{figure*}

The intrinsic luminosity of an AGN is a fundamental property that can be used to estimate  AGN accretion rates, compare multiple populations of AGNs, and explore the link between AGN activity and star-formation in galaxies. In J1816NE and B, we are presented with three different, independent methods to infer the AGN luminosity. The X-ray luminosity traces emission from very near the AGN accretion disk, as photons emitted here are inverse Compton scattered to X-ray energies by the cloud of hot electrons near the accretion disk \citep[e.g.,][]{brandt2005,comastri2008}. Similarly, the mid-IR emission derived from SED modeling is emitted as reprocessed emission from the distribution of dust surrounding and heated by accretion disk photons \citep[e.g.,][]{efstathiou1995, alonsoherrero2001}. The luminosity in the strong [\ion{O}{3}]$\lambda$5007 emission line is also used as an AGN luminosity indicator, and arises from recombinations in the pc- to kpc-scale NLRs photo-ionized by AGNs. There are strong relations between these indicators \citep[see e.g.,][]{lamassa2010}, although these relations may break down in more luminous AGNs \citep[][Chen et al. in prep]{stern2015}, highly obscured AGNs, or in AGNs with multiple regions of extinction \citep{goulding2009}.

In \citet{hainline2013,hainline2014a}, the authors explored the relationship between the physical extent of the NLR and the AGN luminosity as traced by both the [\ion{O}{3}]$\lambda$5007 and mid-IR luminosity indicators, demonstrating that the most powerful AGNs were capable of ionizing gas on scales of up to $\sim 10$ kpc, corresponding to the full extent of the gas in the host galaxy. As has been shown in previous sections, J1816NE has an extended kpc-scale, kinematically complex NLR, while on the other hand J1816SW has a very compact NLR. It is believed that the extended emission observed in J1816NE is due to gas disrupted in the merger being photoionized in the past by the nuclear AGN in J1816NE, and \citet{keel2015} hypothesized that the extended [\ion{O}{3}] emission could not be ionized by the relatively weak nuclear AGN, indicating that this object was a ``fading'' AGN, where it was more luminous in the recent past ($< 10^5$ yrs). Understanding the intrinsic luminosity of an AGN is vital for comparing large AGN samples, and emission from extended NLRs might lead to an overestimate of the AGN luminosity, especially at high-redshift, where they may not be resolved apart from the galaxy nucleus. Thus, J1816NE and J1816SW represent a useful pair to explore how the [\ion{O}{3}] luminosity of an AGN is dependent on the size of the region used to explore the extended emission, especially with respect to fiber-based spectroscopy such as that employed by SDSS. 

\begin{deluxetable*}{llccc}
\tabletypesize{\scriptsize}
\tablecaption{[\ion{O}{3}] Uncorrected Flux, Corrected Flux and Luminosity \label{tab:oiiiflux}}
\tablewidth{0pt}
\tablehead{
\colhead {Object} & \colhead{Aperture} & \colhead{Flux} & \colhead{Flux (corrected)} & \colhead{log(L$_{[OIII]}$)} \\
\colhead{} & \colhead{Size} & \colhead{$10^{-16} \mathrm{erg}\;\mathrm{cm}^{-2}\;\mathrm{s}^{-1}$} & \colhead{$10^{-16} \mathrm{erg} \;\mathrm{cm}^{-2}\;\mathrm{s}^{-1}$} & \colhead{erg s$^{-1}$}
}
\startdata

 & 1\arcsec\ & $255.5 \pm 3.3$ & $182.8 \pm 2.3$ & 40.8 \\
 & 2\arcsec\ & $300.7 \pm 2.6$ & $389.1 \pm 3.4$ & 41.2 \\
J1816NE & 3\arcsec\ & $358.1 \pm 2.9$ & $605.6 \pm 5.0$ & 41.4 \\
 & 5\arcsec\ & $458.6 \pm 3.3$ & $987.8 \pm 7.1$ & 41.6 \\
 & 10\arcsec\ & $593.0 \pm 101.6$ & $1343.8 \pm 230.2$ & 41.7 \\ 
\\
 & 1\arcsec\ & $98.9 \pm 3.6$ & $81.0 \pm 3.0$ & 40.5 \\
 & 2\arcsec\ & $92.2 \pm 13.7$ & $103.4 \pm 15.4$ & 40.6 \\
J1816SW & 3\arcsec\ & $87.1 \pm 5.3$ & $92.1 \pm 5.6$ & 40.5 \\
 & 5\arcsec\ & $82.9 \pm 5.0$ & $121.5 \pm 7.3$ & 40.7 \\
 & 10\arcsec\ & $75.9 \pm 5.7$ & $355.0 \pm 26.5$ & 41.1 
\enddata
\end{deluxetable*}

We used the April 19 long-slit MDM observations of J1816NE, and the June 30 observations of J1816SW to measure the total [\ion{O}{3}] emission in varying extraction regions. We began by extracting spectra in regions corresponding to 1\arcsec, 2\arcsec, 3\arcsec, 5\arcsec, and 10\arcsec\ from the two-dimensional spectrum for each object. We then removed the stellar continuum, and fitted the [\ion{O}{3}] emission line complex in a manner similar to the method used in Section \ref{sec:kinematicsandionization} to measure the total flux of the [\ion{O}{3}] emission line for each of these extracted one-dimensional spectra. To correct for the fact that these measurements were made using long-slit data, and not from circular apertures, we used the existing {\em HST} broad- and ramp-filter images to explore how the observed counts changed as a function of radius. In both images, we measured the counts in multiple slit and circular apertures of increasing size, and then subtracted the counts seen in the apertures in the scaled broad-filter image from the counts seen in the apertures in the ramp-filter image. By comparing these counts for both the slit and circular apertures we estimated correction factors, which were then applied to fluxes measured from the long-slit spectra. We report both the measured fluxes and the corrected fluxes and luminosities in Table \ref{tab:oiiiflux}. 

In the top-left panel of Figure \ref{fig:oiii_vs_ir}, we plot the mid-IR luminosity derived from the SED fitting outlined in \S \ref{sec:broadband} against the [\ion{O}{3}]$\lambda$5007 luminosities measured from this analysis for both J1816NE and J1816SW. As a comparison, we also plot the local AGNs from \citet{lamassa2010}: an ``[\ion{O}{3}] sample,'' 20 Type 2 Seyfert galaxies (19 of the 20 objects are at $z < 0.1$) chosen from a larger SDSS sample of AGNs, the ``12-$\mu$m sample,'' 31 Type 2 Seyfert galaxies ($z < 0.1$), a complete sample down to a flux density limit of 0.3 Jy at 12$\mu$m from the IRAS Point Source Catalogue with latitude $|b| > 25^{\circ}$. These were split into two samples based on strong or weak observed polycyclic aromatic hydrocarbon (PAH) emission in the plot. As can be seen from the figure and from Table \ref{tab:oiiiflux}, the luminosity for J1816NE spans an order of magnitude from the $1''$ to the $10''$ diameter extraction aperture. On the other hand, the [\ion{O}{3}] luminosity for J1816SW, an object with a more compact NLR, does not significantly change with an increase in the aperture size, although at the largest aperture, there is an increase due to faint [\ion{O}{3}] emission that our slit did not probe. Presumably, this emission is [\ion{O}{3}]-emitting gas stripped during the galaxy interaction, as seen in Fabry-Perot imaging of the pair from \citet{keel2015}. 

The top-right and bottom-left panels of Figure \ref{fig:oiii_vs_ir} compare the unabsorbed $2-10$ keV X-ray luminosities for J1816NE and J1816SW to the mid-infrared and [\ion{O}{3}] luminosities, along with the same samples of local AGNs from the literature. In the top-right panel we plot $2-10$ keV X-ray luminosity against [\ion{O}{3}] luminosity, and for comparison we also plot a sample of local Type 1 AGNs from \citet{heckman2005}. We do not include the \citet{heckman2005} Type 2 AGNs in our comparison, as these authors only provide \textit{observed} X-ray luminosities and intervening neutral gas absorption causes a larger spread to smaller $L_{2-10\, \mathrm{keV}}$ values for these objects. As a result, we feel that a more direct comparison can be made between our unabsorbed luminosity and the \citet{heckman2005} Type 1 AGNs. Both J1816NE and J1816SW have luminosities that are similar to local Type 1 comparison sample. For J1816NE, as the aperture used to measure the [\ion{O}{3}] luminosity increases, the measured luminosity better agrees with the locus of the local AGNs. In the bottom-left panel we plot intrinsic $2-10$ keV X-ray luminosity against monochromatic luminosity at 12.3$\mu$m in order to compare to a sample of local Seyfert and Compton-Thick AGNs from \citet{gandhi2009}. Here, the intrinsic X-ray luminosities for both J1816NE and J1816SW agree with the local samples.

\begin{figure*}[ht]
	\epsscale{1.2} 
	\plotone{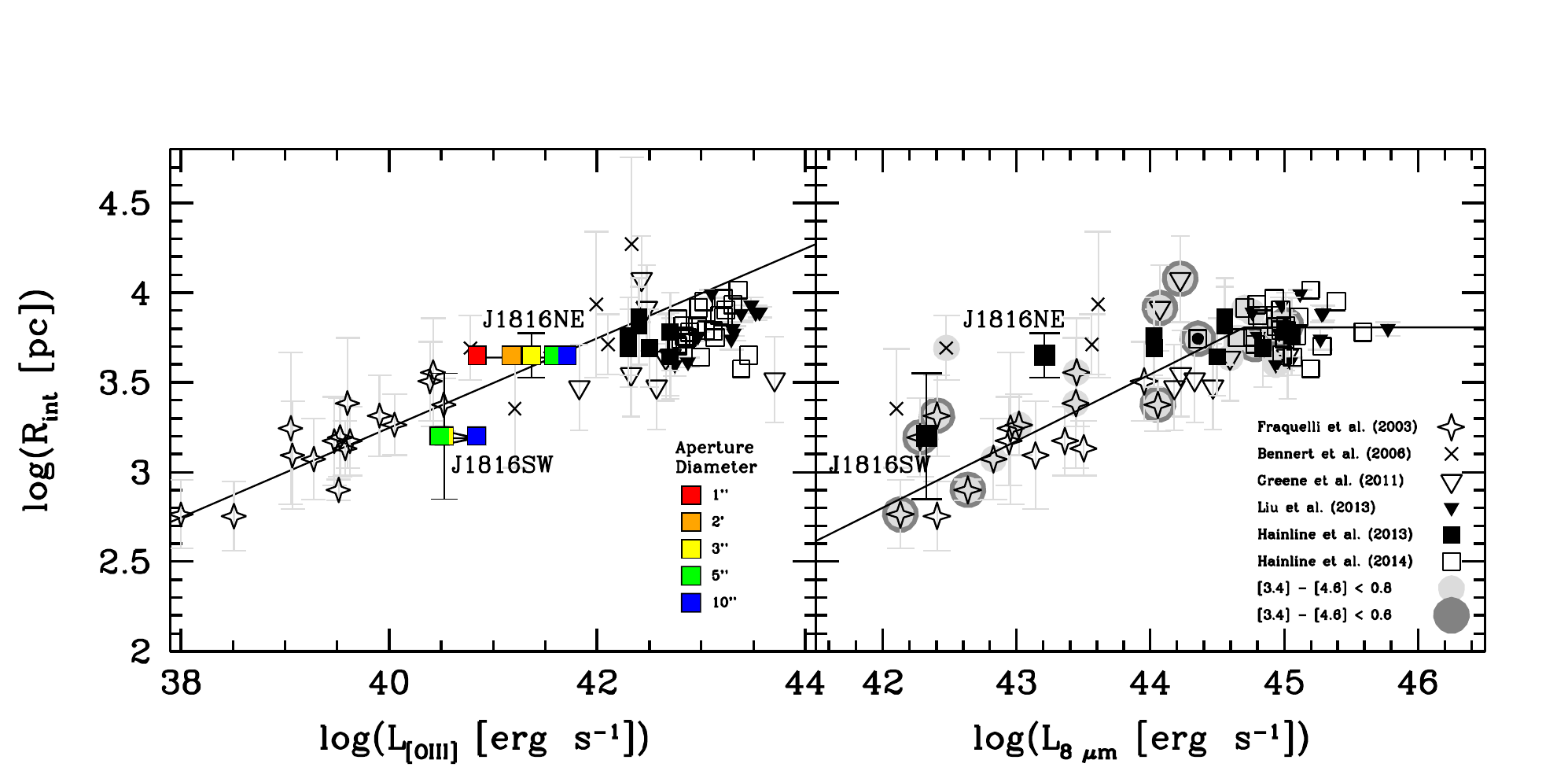}
	\caption{
	\label{fig:lum_vs_rad} Comparison of the estimated size of the NLR in J1816NE and J1816SW with the [\ion{O}{3}] luminosity (left) and 8$\mu$m luminosity (right). NLR sizes are plotted using $R_{int}$, defined as the size of the object at a limiting surface brightness, corrected for cosmological dimming, of $10^{-15}/(1+z)^4$ erg cm$^{-2}$ s$^{-1}$ arcsec$^{-2}$. For comparison, we plot AGNs with sizes from the literature, including Seyfert galaxies from \citet[][open stars]{fraquelli2003} and \citet[][crosses]{bennert2006}, as well as obscured quasars from \citet[][open triangles]{greene2011}, \citet[][black filled triangles]{liu2013}, \citet[][black filled squares]{hainline2013}, and \citet[][black open squares]{hainline2014a}. The \citet{hainline2014a} points are those measured using a S\'{e}rsic profile, and the \citet{greene2011} and \citet{liu2013} sizes were corrected by 0.2 dex, in line with their average seeing. On the right, we highlight those objects with WISE color $[3.4]-[4.6] < 0.8$ superimposed on a light gray circle, and those objects with WISE color $[3.4]-[4.6] < 0.6$ over a dark gray circle to indicate which objects may be suffering from contamination by stellar processes. On the left plot, we show the sizes for J1816NE and J1816SW with squares colored by the size of the aperture used to measure $L_{\mathrm{[OIII]}}$. The black line is a best-fit to the data from \citet{hainline2013}. On the right we show the sizes for J1816NE and J1816SW using squares, and the line is a best-fit to the data from \citet{hainline2014a}. While J1816NE and J1816SW have NLR sizes that agree with objects of similar AGN luminosity, J1816NE has a significantly larger NLR than would be expected from both the mid-IR and the 1\arcsec\ nuclear aperture extraction of [\ion{O}{3}]. } 	 
	\epsscale{1.5}
         \end{figure*}

We can also compare our estimates for the luminosities with the measured sizes of the NLRs in J1816NE and J1816SW. As discussed in \citet{greene2011,liu2013,liu2014,hainline2013,hainline2014a}, the average size of the NLR increases in more luminous AGNs. We can measure the NLR size for J1816NE and J1816SW using $R_{int}$, which is defined as the size of the [\ion{O}{3}]$\lambda$5007 emitting region to a limiting surface brightness of, $10^{-15}/(1+z)^4$ erg cm$^{-2}$ s$^{-1}$ arcsec$^{-2}$, which has been corrected for cosmological dimming. To calculate $R_{int}$, we used the [\ion{O}{3}] fluxes measured from the one dimensional spectra for each extracted aperture along the slit, using the aperture size ($1\secpoint6$) and slit width ($1\secpoint2$) to calculate the surface brightness. We then estimated the radius to which the surface brightness fell to $10^{-15}/(1+z)^4$ erg cm$^{-2}$ s$^{-1}$ arcsec$^{-2}$ for each slit, and our uncertainties were estimated using the size of the apertures that we used to measure the individual spectra, which was dictated by the average seeing for the observations. For J1816NE, we measure $R_{int} = 5.4 \pm 1.3$ kpc from the 126$^\circ$ April 19 slit, and $R_{int} = 4.5 \pm 1.3$ kpc from the 59$^\circ$ April 21 and June 30 slits. For J1816SW, we measure $R_{int} = 1.6 \pm 1.3$ kpc for the 59$^\circ$ and 25$^\circ$ June 30 slits and $R_{int} = 1.3 \pm 1.3$ kpc for the 59$^\circ$ April 21 slit. For the remainder of this section, we will use $R_{int} = 4.5 \pm 1.3$ kpc for J1816NE and $R_{int} = 1.6 \pm 1.3$ kpc for J1816SW, noting that our results do not change significantly if we use the other measured sizes. We show our sizes against the [\ion{O}{3}] and AGN IR luminosity in Figure \ref{fig:lum_vs_rad} along with NLR size measurements from the literature, as well as best-fits to the relationships from \citet{hainline2013,hainline2014a}. On the right side of the figure, we parameterize AGN IR luminosity with $L_{8\mu m}$, the luminosity at 8 $\mu$m estimated from the best-fitting SED without extinction. 

The sizes for J1816NE and J1816SW largely agree with the estimates taken from the literature, although both NLR sizes are larger than would be predicted from the IR luminosity. (We note that if the AGN were responsible for the entirety of the observed WISE emission, the values for the IR luminosity would change to log$(L_{8\mu m} / \mathrm{erg\; s}^{-1}) <  43.3$ for J1816NE and log$(L_{8\mu m} / \mathrm{erg\; s}^{-1}) <  42.8$ for J1816SW, which better agrees with the best-fit.) On the NLR size vs [\ion{O}{3}] luminosity diagram, as the size of the aperture used to measure the [\ion{O}{3}] luminosity increases, J1816NE's estimated NLR size better agrees with the observed relation, which indicates the care that should be taken when using [\ion{O}{3}] as a luminosity proxy in more distant galaxies where the extended NLR is not resolved. 

\section{Discussion and Conclusions}
\label{sec:conclusions}

The kinematic and SED analysis of the J1816 merging system reveals an intriguing pair of very different galaxies. From Figures \ref{fig:sdssimage} and \ref{fig:multiwaveimage}, it can be seen that J1816NE is the more morphologically disturbed of the two merging galaxies. In J1816NE, the more luminous AGN, gas is illuminated out to $>5$ kpc from the nucleus in a clear bicone and a velocity gradient of $\sim 500$ km s$^{-1}$, with a velocity dispersion of around $> 200-300$ km s$^{-1}$ near the nucleus. This gas has emission line ratios indicative of AGN ionization, with high [\ion{O}{3}] / H$\beta$ ratios at a large radius from the galaxy nucleus. In contrast, J1816 has a much more compact (1-2 kpc in radius) and less kinematically disturbed emission line region. There is no strong gradient across the region, and the velocity dispersion is only $\sim 100$ km s$^{-1}$. The ionization in the J1816SW NLR is as strong as what is observed across the large NLR in J1816NE. 

SED decomposition reveals similar host galaxy star-formation properties ($2-3$ M$_{\sun}$ yr$^{-1}$) for the pair, but with strikingly different AGN properties. The AGN in J1816NE is significantly more luminous than the AGN in J1816SW, which is the more obscured of the pair, in agreement with the results comparing both the integrated [\ion{O}{3}] and observed X-ray luminosities. When we directly compare the luminosities of J1816NE and J1816SW with local obscured and unobscured AGNs, a clear picture emerges. J1816SW is a lower luminosity obscured AGN whose infrared, [\ion{O}{3}], and intrinsic 2-10 keV X-ray luminosities agree with local samples. J1816NE, on the other hand, has a large intrinsic X-ray luminosity, which correlates with the [\ion{O}{3}] luminosity only when measured across the full extended emission line region. This result highlights the potential importance of aperture effects in using extended ionized gas emission as a bolometric luminosity indicator, and indicates that, for J1816NE, the luminosity of the extended [\ion{O}{3}] emission reflects the current level of nuclear activity. In addition, for J1816NE, the kinematic size of the NLR, i.e., the radius at which the velocity of the ionized gas in the extended region drops to the systemic velocity, agrees with what is estimated from the surface brightness of the region. 

Based on the existence of extended tidal features seen in the optical continuum images seen in Figures \ref{fig:sdssimage} and \ref{fig:multiwaveimage}, and an ionized tidal bridge between the two galaxies (Figure \ref{fig:twodonedspectra}), it is likely that the J1816 merger pair has undergone one close encounter in the past. The merger has clearly more strongly affected J1816NE, which has an irregular optical morphology and a large quantity of extended gas, than J1816SW, which still retains a disk morphology. From the timescales derived for simulated mergers \citep[e.g.,][]{dimatteo2005,cox2008,hopkins2008}, it is likely that we are seeing the J1816 merger after an interaction of over a Gyr. We are currently observing a large quantity of ionized material moving away from the nucleus of J1816NE, as observed in the outflowing gas in the ionization cones. Meanwhile, a more  extended and kinematically distinct component of the ionized gas is coincident with extended tidal features in the stellar continuum. The tidal torques will correspondingly drive gas towards the center of the galaxy and may help fuel star formation and black hole accretion. In the case of J1816NE, we appear to be observing both AGN accretion and feedback, and it is possible that J1816NE may eventually be powerful enough to cease the infall of material onto the black hole. As the merger progresses, J1816SW, then, may go through a similar increase in AGN luminosity as more gas is fed to the black hole. 

While the J1816NE and J1816SW pair has a larger separation, the major merger of the two galaxies may result in a dual AGN state in the future \citep[e.g.,][]{komossa2003, koss2011, mazzarella2012}. Recently, \citet{comerford2015} explored the hard X-ray and [\ion{O}{3}] luminosity for a sample of ``dual AGNs,'' galaxies observed with two AGNs separated by less than 10 kpc. For their sample, they found that the observed hard X-ray luminosities were, on average, half what would be observed for single AGNs at a given [\ion{O}{3}]$\lambda$5007 luminosity. The \citet{comerford2015} results then imply a strong evolution in the X-ray luminosity for these objects, or at least an increase in the dust obscuration that may lead to a systematic lowering of the X-ray luminosity from what is observed in Figure \ref{fig:oiii_vs_ir}. Understanding the evolution of dual AGN phases from wide separation after first encounter (such as in the J1816 system) to a binary phase and finally coalescence provides interesting insights into the cosmological growth of black holes.

As larger photometric and spectroscopic samples of active galaxies are explored, more objects with extended NLRs will be uncovered. The upcoming SDSS-IV MaNGA Survey \citep{bundy2015} will result in IFU observations for 10,000 nearby galaxies, including many AGNs. These data will allow for a detailed analysis of the gas ionization and kinematics across the face of these galaxies similar to what we have done with J1816NE and J1816SW. Importantly, while extended NLRs are ubiquitous at quasar luminosities, these future deep observations should probe lower intrinsic AGN luminosities and explore the properties of a large sample of extended NLRs as a function of AGN luminosity. These objects, which give us an understanding of current and past AGN activity, are important for placing AGNs in the context of galaxy evolution. 

\acknowledgments 

KNH and RCH were partially supported by NASA through ADAP award NNX12AE38G and by the National Science Foundation through grant numbers 1211096 and 1211112. RCH acknowledges support from an Alfred P. Sloan Research Fellowship and a Dartmouth Class of 1962 Faculty Fellowship. The authors would like to thank Bill Keel, whose input was helpful in preparing this manuscript. This work is based on observations obtained at the MDM Observatory, operated by Dartmouth College, Columbia University, Ohio State University, Ohio University, and the University of Michigan. This work also utilizes {\it Herschel} data. {\it Herschel} is an ESA space observatory with science instruments provided by European-led Principal Investigator consortia and with important participation from NASA. Finally, this publication makes use of data products from the Wide-field Infrared Survey Explorer, which is a joint project of the University of California, Los Angeles, and the Jet Propulsion Laboratory/California Institute of Technology, and NEOWISE, which is a project of the Jet Propulsion Laboratory/California Institute of Technology. WISE and NEOWISE are funded by the National Aeronautics and Space Administration.

\bibliographystyle{apj}

\end{document}